\documentclass[11pt]{article}

\usepackage{graphicx}
\usepackage{amsmath,amssymb}
\usepackage{amsmath}
\usepackage{xcolor}
\usepackage{caption}
\usepackage{float}
\usepackage{ragged2e}
\usepackage{subcaption}
\usepackage[numbers]{natbib}

\usepackage[body={6in,8.5in}]{geometry}

\title{WHOI Proceedings 2024 \\ Oceanic internal tides: do they get phased at the Equator?}
\author{Camille Moisset, Bruce Sutherland, Lois Baker}
\date{\today}

\begin{document}

\maketitle

% LEAVE THESE NEXT TWO LINES IN (used in final version for page numbering).
% They must go right after the \maketitle command.
%\setcounter{page}{\pageref{pr-LastPage}}
%\addtocounter{page}{1}

\section{Introduction}\label{intro}
Oceanic tides are an essential component of the global ocean circulation, with an important role in the exchanges between the different energy reservoirs \citep[][]{MunkWunsch1998, FerrariWunsch2009}. However, two different contributions need to be separated: the first is the barotropic tide, generated by astronomical forcing, which produces a large scale movement of water masses across the entire global ocean; the second is baroclinic tide (or internal tide) which are internal waves excited by the movement of the water masses over topography \citep[e.g.][]{Gerkema2008, Sutherland2010}. These waves then propagate away from their sources as low vertical modes and interact with numerous other oceanic processes \citep[e.g.][]{Mackinnonetal2017}. The dynamics of internal tides is incredibly rich involving complex generation mechanisms, their interactions with topography, but also wave-wave and wave-mean flow interactions, all through the stratified and rotating fluid that is the ocean \cite[see][for reviews on different dynamical aspects]{GarrettKunze2007, Lamb2014, SarkarScotti2017}. Internal tides are a significant source for deep oceanic mixing through their non-linear dynamics that ultimately induce turbulent dissipation and diapycnal mixing.\\

The lunar semidiurnal internal tide, denoted M2, is the most energetic tidal mode excited in Earth oceans and, as such, a major actor of oceanic dynamics. It is excited at a fixed frequency $\omega_{0}=1.4\times10^{-4}$ rad.s$^{-1}$ and propagates over long distances as low modes with a horizontal wavelength around 150 kilometers \citep[e.g.][]{Martinetal2006, deLavergneetal2019}. The ultimate fate of the baroclinic tides remains uncertain but a significant number of processes are proposed, amongst them non-linear breaking, triadic interactions, scattering of the energy to higher modes, and interaction with currents and eddies. Likewise, the interaction between the internal tides and eddy fields could cause such incoherences in the tidal signal \citep[][]{RainvillePinkel2004}. It is also possible that strong currents and shelf fronts influence the dynamics of the M2 internal tide \citep[][]{KellyLermusiaux2016}.\\

\begin{figure}
    \centering
    \includegraphics[width=0.6\linewidth]{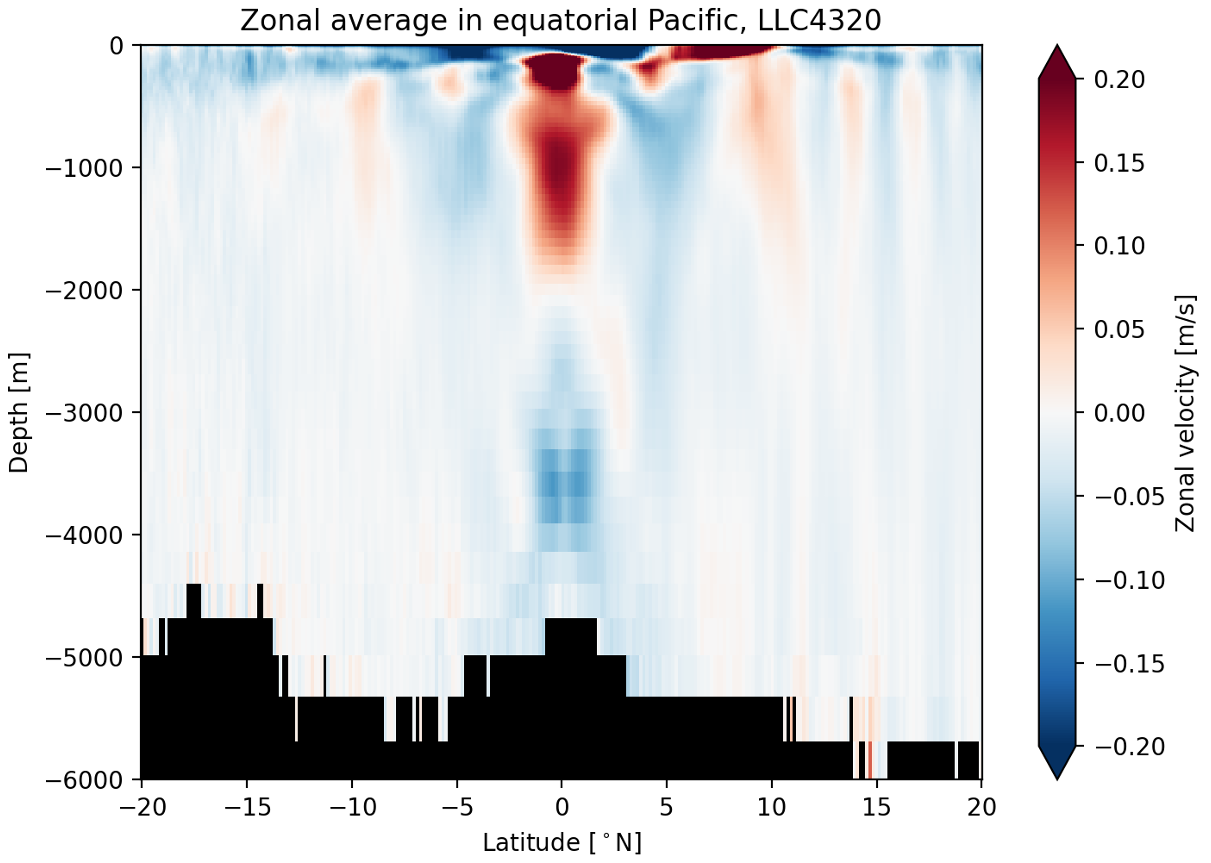}
    \captionsetup{width=1.\linewidth, justification=justified, format=plain}
    \caption{Vertical structure of the zonal equatorial current, computed using the LLC4320 configuration of the Massachusetts Institute of Technology general circulation model (MITgcm, \citep[][]{Marshalletal1997})}
    \label{presentation_equatorial_jet}
\end{figure}

Satellite altimetry provides a way to track the baroclinic tidal signal over the global ocean, as oscillations of the interior affect the surface of the ocean. But in the region of the Equatorial Pacific, such observations show that the M2 signal (for example propagating northward from its generation near the French Polynesia Islands) seems to disappear when the internal waves cross the Equator (see fig. 1 from \citep[][]{Buijsmanetal2017}). Buijsman et al. \citep[][]{Buijsmanetal2017} argue that it could be due to a loss of coherence of the M2 signal and they hinted that equatorial jets (see figure \ref{presentation_equatorial_jet}) could play a significant role: the vertical and meridional variations in the zonal currents and the stratification can scatter the energy to higher modes and modify the phase and group speed of internal waves resulting in their incoherences. In such type of observations, the signal of the M2 internal tide is extracted using a harmonic fit to the excitation frequency of the tidal mode, meaning that incoherences could induce a significant loss of information.\\

\noindent We investigate whether the dynamics of the wave-mean flow interaction between the M2 internal wave and the equatorial jets could cause a loss of coherence in the M2 altimetry signal. To do so, we study the more general interaction between a vertical mode 1 internal tide wavepacket and a vertically sheared zonal jet. We follow the work of Kelly et al. \citep[][]{Kelly2016, Kellyetal2016, Kellyetal2021} who proposed a model able to decompose the modal dynamics of the M2 internal tides. In particular, linear wave-mean flow interactions are retained but non-linear wave-wave interactions are neglected. Section \ref{analytics} details the formalism of the vertical mode decomposition for internal tides, as well as the analytical framework used in our model. Simulations are realised using Dedalus \citep[][]{Burnsetal2020}, the numerical setup and parameters which are presented in section \ref{num param}. Section \ref{toymodel} show the results for a simplified model in which the ocean is supposed to linearly stratified. This approximation allows for the exact comparison between the analytic calculations and the simulations. The conclusions and perspectives for future work are presented in section \ref{conclusion}.

\section{Analytical framework}\label{analytics}
\subsection{General equations}
We use the tangent plane approximation to describe a patch of the Equatorial Pacific Ocean as a Cartesian box, where $(x,y,z)$ represent longitude, latitude and depth respectively, with unit vectors $(\boldsymbol{e_x},\boldsymbol{e_y}, \boldsymbol{e_z})$. We consider the fully non-linear Euler equations with the Boussinesq approximation:
\begin{subequations}
\begin{align}
    \displaystyle \partial_t \boldsymbol{u_{T}}+(\boldsymbol{u_{T}}\cdot\boldsymbol{\nabla})\boldsymbol{u_{T}}+\boldsymbol{f}\times\boldsymbol{u_{T}} &=-\boldsymbol{\nabla} p+b_{T}\boldsymbol{e_{z}},\\
    \displaystyle \partial_t b_{T}+\boldsymbol{u_{T}}\cdot\boldsymbol{\nabla}b_{T}&=0,\\
    \displaystyle \boldsymbol{\nabla}\cdot\boldsymbol{u_{T}}&=0,
\end{align}
\label{systeme_vectoriel}
\end{subequations}
\noindent where $\boldsymbol{u_{T}}=(u_{T}, v_{T}, w_{T})=(U+u,v,w)$ is the total velocity field (including currents and waves, with $U$ the velocity of the current), $\boldsymbol{f}=2\bf{\Omega_{0}}\,\rm{sin}\lambda$ the Coriolis parameter with $\bf{\Omega_{0}}$ the rotation vector and $\lambda$ the latitude. $p=P/\rho_0$ is the reduced pressure where $P$ is the dynamic pressure, the density has been decomposed as $\rho=\rho_0+\overline{\rho}(z)+\rho(x,y,z,t)$ with $\rho_0$ a reference density, $\overline{\rho}$ the mean density profile and $\rho$ the perturbation density. The squared buoyancy frequency is $\displaystyle \overline{N}^{2}=-g\overline{\rho}'/\rho_{0}=-\overline{b}'$, where $g$ is gravity and the primes denote derivatives. The total buoyancy field is $b_{T}=B+\overline{b}+b$ with $B$ the buoyancy perturbation corresponding to the zonal current, $\overline{b}$ that of the background stratification and $b$ that of the waves.\\

At the Equator, in the $f$-plane approximation the traditional Coriolis parameter $\boldsymbol{f}=(0,0,f)$ goes to zero so we use the $\beta$-plane approximation $f = \beta y$ (with constant $\beta=2\times10^{-11}$ m$^{-1}$s$^{-1}$).\\

\noindent The mean flow $U$ and its associated buoyancy $B$ and pressure fields $P_{U}$ verify the geostrophic and hydrostatic balance:
\begin{subequations}
\begin{align}
	\beta y U&=-\partial_{y} P_{U},\\
	0&=-\partial_{z} P_{U} + B,
\end{align}
\label{systeme_mean_flow}
\end{subequations}
which leads to
\begin{equation}
	\partial_{y}B=-\beta y \partial_{z} U.
	\label{derivative_B_definition}
\end{equation}

\subsection{Base state}
The base velocity considered is a zonal jet with separated meridional and vertical dependencies:
\begin{equation}
    \boldsymbol{U}=(U(y,z),0,0) \quad \textrm{with} \quad
    U(y,z)=E(y)\tilde{U}(z),
    \label{base_velocity}
\end{equation}
where the meridional structure $E(y)$ is taken as a Gaussian:
\begin{equation}
    E(y) = \textrm{exp}\left(-\frac{y^2}{2\rm{W}^2}\right)\,,
\end{equation}
with the characteristic width of the jet denoted by $\rm{W}$.\\
In what follows, $\tilde{U}(z)$ will be either a constant or a cosine function and $\overline{N}^2$ will be a constant. It is however possible to generalise to more complex and realistic jet configurations and background stratification.\\

\noindent Using (\ref{derivative_B_definition}), the associated base buoyancy is in hydrostatic and equatorial geostrophic balance:
\begin{equation}
    B(y,z) = \beta\, \textrm{W}^2 E(y)\tilde{U}'(z).
    \label{base_buoyancy}
\end{equation}

\noindent The background stratification includes the contribution of the Equatorial jet $(\partial_{z}B)$ and the far field stratification:
\begin{equation}
    N^2(y,z) = \beta \textrm{W}^2E(y)\tilde{U}''(z) + \overline{N}^2(z)\,.
\end{equation}
\subsection{Equations for the perturbations}
The base state (\ref{base_velocity}-\ref{base_buoyancy}) is subjected to perturbation by the internal tide. The total fields are given by:
\begin{equation}
    (u_{\textrm{T}},v_{\textrm{T}},w_{\textrm{T}},p_{\textrm{T}},b_{\textrm{T}})=(U,0,0,P_{U},B)+(u, v, w, p, b).
\end{equation}
Substituting and neglecting non-linear wave-wave interaction terms in (\ref{systeme_vectoriel}) yields the linearized system for the evolution of waves:
\begin{subequations}
    \begin{align}
    \displaystyle \partial_t u + E\tilde{U}\partial_x u + E'\tilde{U}v + E\tilde{U}'w - \beta y v &= -\partial_x p, \label{umomlinear}\\
    \displaystyle \partial_t v + E\tilde{U}\partial_x v + \beta y u &= -\partial_y p, \label{vmomlinear}\\
    \displaystyle \alpha \left(\partial_t w + E\tilde{U}\partial_x w\right) &= -\partial_z p + b, \label{wmomlinear}\\
    \displaystyle \partial_t b + E\tilde{U}\partial_x b + E'\tilde{U}' \beta \textrm{W}^2 v + \beta \textrm{W}^2E\tilde{U}''w + \overline{N}^2w &= 0,\label{buoylinear}\\
    \displaystyle \partial_x u+\partial_y v+\partial_z w &= 0 \label{incomp}
    \end{align}
    \label{system_perturbations}
\end{subequations}
where $\alpha \in \{0,1\}$ allows to distinguish between the hydrostatic approximation ($\alpha = 0$) and non-hydrostatic case ($\alpha = 1$).

\subsection{Vertical mode decomposition for internal tides}
We will consider internal tides incident upon and interacting with equatorial jets. They are initialised far from the equator, where the equatorial currents are negligibly small. Here we describe the process of determining the internal structure of these waves.
\subsubsection{General formalism}\label{sect_wave_general}
We derive the vertical structure of internal tides with no background flow $(U=0)$. Hydrostatic balance is also assumed by setting $\alpha$ to zero. The system (\ref{system_perturbations}) is linear and we expand the perturbations using the plane wave ansatz:
\begin{equation}
    (u, v, w, p, b)(x,y,z,t) = (\hat{u},\hat{v},\hat{w},\hat{p},\hat{b}) (z) e^{i(k_{x}x + k_{y}y - \omega t)}\,,
    \label{wave_ansatz}
\end{equation}
where $k_{x}$ and $k_{y}$ are the wavenumbers in the zonal and meridional directions respectively and $\omega$ is the frequency. The system (\ref{system_perturbations}) becomes:
\begin{subequations}
\begin{align}
    -i\omega \hat u - \beta y_{0} \hat{v} &= - ik_{x} \hat p, \label{system_waves_a} \\
    -i \omega v + \beta y_{0} \hat{u} & = -ik_{y} \hat p, \label{system_waves_b} \\
    0 &= -\hat p' + \hat b, \label{system_waves_c} \\
    -i \omega \hat b + \overline{N}^2\hat w &= 0, \label{system_waves_d} \\
    ik_{x} \hat u + i k_{y} \hat v + \hat w' &= 0\,, \label{system_waves_e}
\end{align}
\label{system_waves}
\end{subequations}
in which $y_{0}$ is the initial latitude of the waves with $|y_{0}|\gg$W.
\noindent The system (\ref{system_waves}) reduces to:
\begin{equation}
    \hat w '' + k_h^2\frac{\overline{N}^2(z)}{\omega^2-(\beta y_{0})^2 } \hat w = 0\,,
\label{weqn}
\end{equation}
where $k_h^2 = k_{x}^2 + k_{y}^2$.\\ 
\noindent The solution of \eqref{weqn} gives a sum of vertical modes $\hat{w}=\Sigma_{n}w_{n}\Phi_{n}(z)$ where the $\Phi_{n}$ satisfy the Sturm-Liouville problem:
\begin{equation}\label{Phieqn}
    \Phi_n'' + \frac{\overline{N}^2(z)}{c_n^2}\Phi_n= 0\,\quad \textrm{with}\quad \Phi_n(0) = \Phi_n(-H) = 0,
\end{equation}
with $c_n^2=(\omega^2-(\beta y_{0})^2)/k_h^2$ as the eigenvalue and $H$ the height of the domain. This gives the dispersion relation $\omega^{2}=c_{n}^{2}k_{h}^{2}+(\beta y_{0})^{2}$. The vertical modes, $\Phi_{n}$, are orthogonal with respect to the weight function $\overline{N}^2(z)$. The normalisation is chosen so that max($|\Phi_n|=1$). Using (\ref{system_waves}) we can derive the corresponding expressions for horizontal velocity, pressure and buoyancy as a sum of vertical modes:
\begin{subequations}
\begin{align}
    [\hat{u},\hat{v},\hat{p}](z) &= \sum_{n=0}^\infty [u_n, v_n, p_n]\phi_n(z),\\
    \hat{b}(z) &= \sum_{n=0}^\infty b_n  \overline{N}^2(z)\Phi_n(z)\,
\end{align}
\label{modal_expansion}
\end{subequations}
where $\phi_{n}=\Phi_{n}'$.

\subsubsection{Case $\overline{N} =$ constant}\label{sect_wave_N_cte}
In the case where $\overline{N}=\overline{N_0}=\rm{constant}$, equation (\ref{Phieqn}) reduces to that of a harmonic oscillator:
\begin{equation}
    \Phi_n'' + m_{n}^2\Phi_n= 0\,\quad \textrm{with}\quad m_{n}^2=\frac{\overline{N}_0^2}{\omega^2-(\beta y_{0})^2}k_h^2.
    \label{SL_pbm}
\end{equation}
Combined with the boundary conditions (\ref{Phieqn}), the vertical structure functions $\Phi_n$ can be expanded as a series of sine functions:
\begin{equation}
    \Phi_n(z)=\textrm{sin}\, (m_{n}z),
\end{equation}
where $m_{n}=n\pi/H$, $n=1,2,...$.\\
Likewise,
\begin{equation}
    \phi_n(z)=m_{n}\textrm{cos}\, (m_{n}z).
\end{equation}

\begin{figure}
    \centering
    \includegraphics[width=0.5\linewidth]{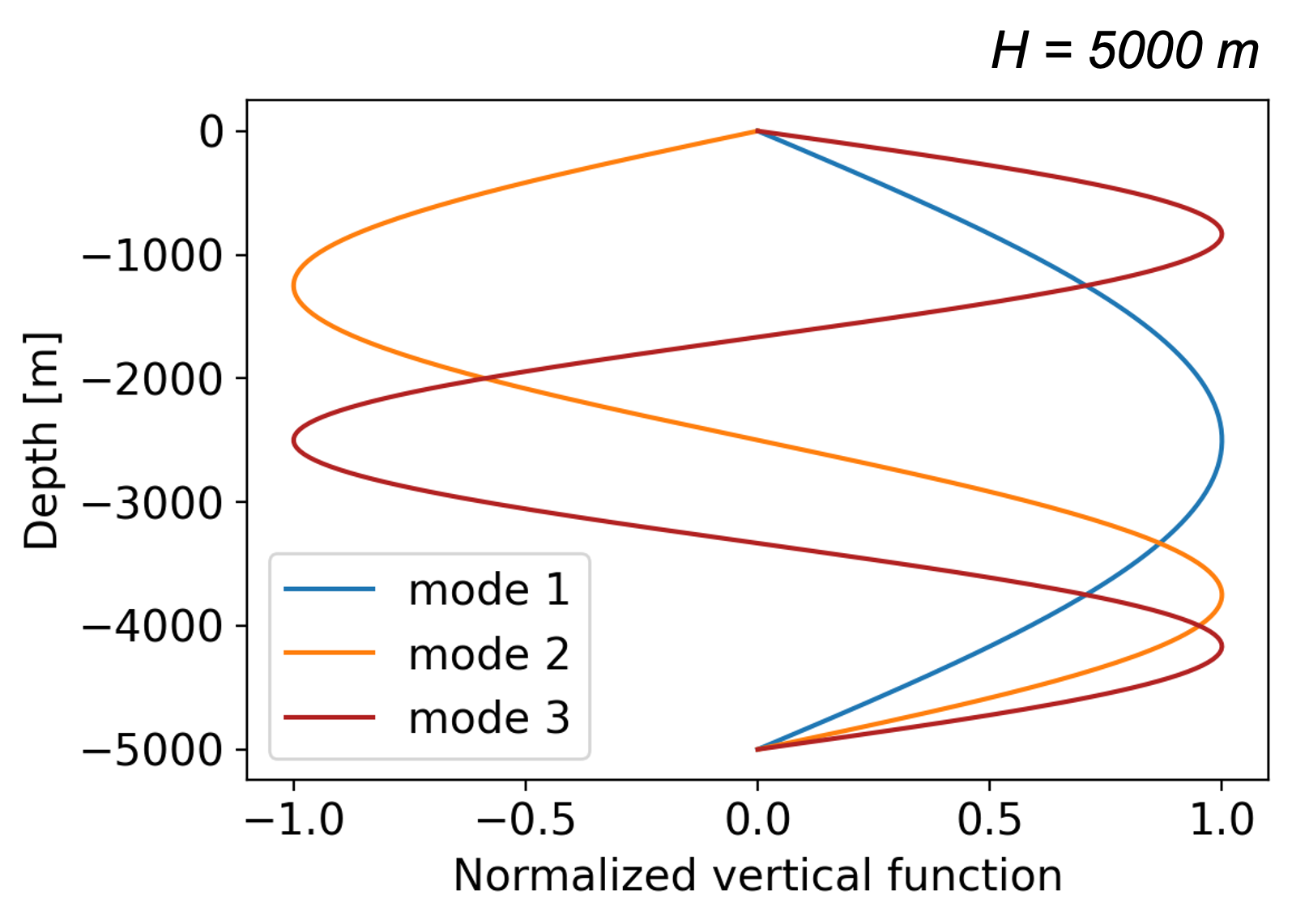}
    \captionsetup{width=1.\linewidth, justification=justified, format=plain}
    \caption{Three first vertical structure functions, $\Phi_{n}$, for $\overline{N_0}^{2}=1\times10^{-4}$ s$^{-2}$ and H$=5000$m.}
    \label{presentation_context}
\end{figure}

\noindent Figure (\ref{presentation_context}) shows the first three vertical structure functions for the case $\overline{N_0}^{2}=1\times10^{-4}$ s$^{-2}$, with H$=5000$m.\\

\noindent From (\ref{SL_pbm}), the dispersion relation is given by
\begin{equation}
    \omega^2=\overline{N_0}^2\left(\frac{k_h}{m_{n}}\right)^2+(\beta y_{0})^2.
\label{relation_dispersion_toymodel}
\end{equation}
Thus, specifying the frequency $\omega$ and mode number $n$ of the initial waves gives the corresponding magnitude of the horizontal wavenumber $k_{h}$.
\subsection{Ray tracing theory}\label{ray_tracing_general}
Ray tracing predicts the path of the wave energy, as it moves at the group velocity $\boldsymbol{c_{g}}$. For horizontal motion:
\begin{equation}
	\boldsymbol{c_{g}}=\boldsymbol{\nabla}_{\boldsymbol{k}}\,\omega=\left(\frac{\partial \omega}{\partial k_{x}},\frac{\partial \omega}{\partial k_{y}}\right)
	\label{cg_def}
\end{equation}
\noindent This is different from the phase velocity $\boldsymbol{c_{p}}$ that gives the velocity of the crests:
\begin{equation}
	\boldsymbol{c_{p}}=\left(\frac{\omega}{|\boldsymbol{k}|}\right)\boldsymbol{\hat{k}}=\frac{\boldsymbol{k}}{|\boldsymbol{k}|^{2}}\omega,
\end{equation}

\noindent In the long wave limit to be considered here, the group and phase velocity are in the same direction.\\

\noindent The addition of a mean flow $\boldsymbol{U}=U(y,z)\boldsymbol{e_{x}}$ will introduce a difference between the absolute frequency $\omega_{0}$ and the intrinsic frequency $\omega_{\textrm{int}}$, that are linked through:
\begin{equation}
	\omega_{0}=\omega_{\textrm{int}}+Uk_{x},
\label{def_abs_frequency}
\end{equation}
where $\omega_{0}=1.4\times10^{-4}$ rad.s$^{-1}$ is the excitation frequency of the M2 internal tide and $\omega_{\textrm{int}}$ the intrinsic frequency satisfies the dispersion relation for waves in stationary fluid. In all the expressions of sections \ref{sect_wave_general} and \ref{sect_wave_N_cte} we replace $\omega\rightarrow\omega_{\textrm{int}}$.\\

\noindent In the absence of mean flow, (\ref{def_abs_frequency}) reduces to:
\begin{equation}
	\omega_{0}=\omega_{\textrm{int}},
\label{simplification_no_flow_abs_frequency}
\end{equation}
and $\omega_{0}$ verifies the dispersion relation and all the results of sections \ref{sect_wave_general} and \ref{sect_wave_N_cte}.\\

\noindent As waves move inside the equatorial jet, it is necessary to separate:
\begin{itemize}
	\item the absolute frequency $\omega_{0}$, which is the frequency observed from the fixed frame: it is composed of the intrinsic frequency $\omega_{\textrm{int}}$ and a Doppler-shift that arise from the movement of the background fluid $U$;
	\item the intrinsic frequency $\omega_{\textrm{int}}$, which is the frequency that corresponds to the frame moving with the mean-flow.
\end{itemize}

\noindent The separation between $\omega_{0}$ and $\omega_{\textrm{int}}$ induced by the presence of the mean flow is inherited by the group velocity (\ref{cg_def}), such that:
\begin{equation}
	c_{gx_{0}}=c_{gx_{\textrm{int}}}+U,\qquad c_{gy_{0}}=c_{gy_{\textrm{int}}}
	\label{group_velocity_abs_mean_flow}
\end{equation}
where $(c_{gx_{\textrm{int}}}, c_{gy_{\textrm{int}}})$ are given by injecting the dispersion relation (\ref{relation_dispersion_toymodel}) in equation (\ref{cg_def}).\\

\noindent For ray tracing to be valid outside and inside of the jet, the vertical and horizontal scale of variations of the waves must be much shorter than the characteristic scale of background variations of $N$ and $U$ respectively. The ray tracing equations yield the evolution of the position of the wavepacket $(x,y)$ and its direction $(k_{x}, k_{y})$:
\begin{subequations}
\begin{align}
    \frac{\textrm{d}x}{\textrm{dt}}&=\frac{\partial \omega_{0}}{\partial k_x}=c_{gx_{\textrm{int}}}+U, \quad\quad \frac{\textrm{d}k_x}{dt}=-\frac{\partial \omega_{0}}{\partial x}, \label{ray_tracing_no_jet_a} \\
    \frac{\textrm{d}y}{\textrm{dt}}&=\frac{\partial \omega_{0}}{\partial k_y}=c_{gy_{\textrm{int}}}, \quad\quad\quad\quad \frac{\textrm{d}k_y}{dt}=-\frac{\partial \omega_{0}}{\partial y}. \label{ray_tracing_no_jet_b}
\end{align}
\label{ray_tracing_no_jet}
\end{subequations}
For both cases (\ref{def_abs_frequency}) or (\ref{simplification_no_flow_abs_frequency}), the dispersion relation yields:
\begin{equation}
    \omega=\omega(k_x, k_y; y)=\omega_{\textrm{int}}+Uk_{x},\quad \omega_{\textrm{int}}=c_{n}^{2}k_{h}^{2}+(\beta y)^{2}.
    \label{ray_tracing_dispersion_simplified}
\end{equation}
which, combined with the right-equation of (\ref{ray_tracing_no_jet_a}) gives
\begin{equation}
    k_x=k_{x_0}=\textrm{constant}.
\end{equation}
The left hand equations of (\ref{ray_tracing_no_jet}), gives the slope in the horizontal of the path:
\begin{equation}
    \frac{\textrm{d}x}{\textrm{d}y}=\frac{c_{gx_{\textrm{int}}}+U}{c_{gy_{\textrm{int}}}}
    \label{result_ray_tracing_nojet}
\end{equation}
Given the dispersion relation (\ref{ray_tracing_dispersion_simplified}), $k_{y}$ is implicitly a function of $y$, likewise $c_{gx}=\partial\omega_{0}/\partial k_{x}$ and $c_{gy}=\partial\omega_{0}/\partial k_{y}$ are implicitly functions of $y$.\\

\noindent If $U=E(y)$, independent of $z$, then (\ref{result_ray_tracing_nojet}) can be integrated to give the path starting at $(x_{0}, y_{0})$:
\begin{equation}
	x(y)=\int_{y_{0}}^{y}\frac{c_{gx_{\textrm{int}}}(\tilde{y})+E(\tilde{y})}{c_{gy_{\textrm{int}}}(\tilde{y})}\textrm{d}\tilde{y}+x_{0}
	\label{ray_tracing_traj_formula}
\end{equation}
\begin{figure}
    \centering
    \includegraphics[width=0.8\linewidth]{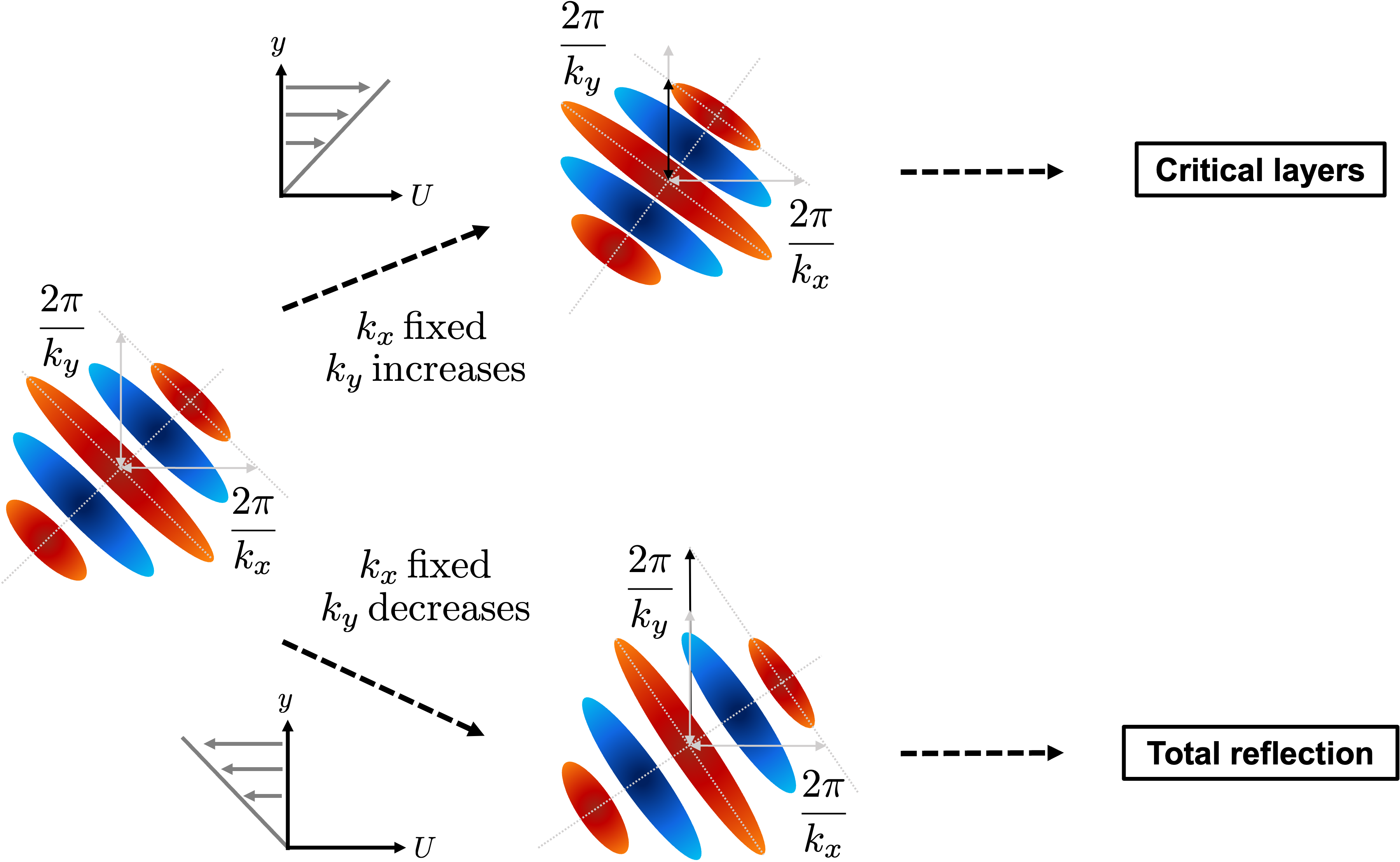}
    \captionsetup{width=\linewidth, justification=justified, format=plain}
    \caption{Schematics of the evolution of the wavepacket when interacting with a uniform jet ignoring the $\beta$ effect. Initial situation is displayed in the center, with wavenumbers drawn in grey arrows. In the upper part, the wavepacket is incident upon a positive horizontal shear so that $k_y$ increases while $k_x$ stays fixed (black arrows), ultimately reacting to a critical layer where $k_{y}\rightarrow\infty$. In the lower part, the wavepacket is incident upon a negative horizontal shear and $k_y$ decreases while $k_x$ stays fixed (black arrows), ultimately leading to a total reflection of the wavepacket where $k_{y}\rightarrow 0$.}
    \label{mechanism_critical_layers_reflexion_uniformjet}
\end{figure}
Ray tracing helps reveal how the $\beta$-plane and jet influence the dynamics of the wavepacket.\\

For example, ignoring the $\beta$-effect and supposing $U=E(y)$, (\ref{ray_tracing_no_jet_b}) yields
\begin{equation}
	\frac{\textrm{d}k_y}{dt}=-k_{x_{0}}E'(y).
\end{equation}
 Figure \ref{mechanism_critical_layers_reflexion_uniformjet} shows how the results from this translate to wave dynamics. Variations of $U$ change $k_{y}$ affecting the wavelength and orientation of the wavepacket. If $E'>0$, then $k_{y}$ decreases. The wavepacket would contract and deviate away from northward (figure \ref{mechanism_critical_layers_reflexion_uniformjet}, bottom right). Ultimately, this could lead to total reflection. Conversely, if $E'<0$ and $k_{y}$ increases, the wavepacket would redirect more toward the north where $k_{y}\rightarrow\infty$ (figure \ref{mechanism_critical_layers_reflexion_uniformjet}, top panel) and this could encounter critical layers.

\subsection{Modal wave equations}
The modal wave equations are obtained by substituting the expansions (\ref{modal_expansion}) into the general equations (\ref{system_perturbations}). Multiplying \eqref{umomlinear}, \eqref{vmomlinear} and \eqref{incomp}  by $\phi_n$, and \eqref{wmomlinear} and \eqref{buoylinear} by $\Phi_n$ and integrating in the vertical, orthogonality allows us to extract the equations for the evolution of the amplitudes of each mode. Equation (\ref{wmomlinear}) gives
\begin{equation}
0 = p_n + c_n^2 b_n\,,
\label{modal_eq_w}
\end{equation}
and equation (\ref{incomp}) gives
\begin{equation}
    \partial_x u_{n} + \partial_y v_{n} + w_n = 0.
    \label{modal_eq_conti}
\end{equation}
These diagnostic equations are used to replace $w_n$ and $p_n$ in the remaining equations for horizontal velocity and buoyancy. From (\ref{umomlinear}):
\begin{equation}
    \partial_tu_{n} + \sum_{m=0}^\infty \left( E\partial_x u_{m} C_{1_{mn}} + v_m E' C_{1_{mn}} + E'(y)w_m C_{2_{mn}}\right) - \beta y v_n - c_n^2 \partial_x b_n= 0,
    \label{modal_eq_u}
\end{equation}
where the interaction coefficients between the vertical structure of the modes and the vertical structure of the jet are given by
\begin{equation}
    C_{1_{nm}}=\displaystyle\frac{\displaystyle\int_{-H}^0\tilde{U}\Phi_n'\Phi_m'\textrm{d}z}{\displaystyle\int_{-H}^0(\Phi_n')^2\textrm{d}z} \quad \textrm{and} \quad C_{2_{nm}}=\displaystyle\frac{\displaystyle\int_{-H}^0\tilde{U}'\Phi_n'\Phi_m\textrm{d}z}{\displaystyle\int_{-H}^0(\Phi_n')^2\textrm{d}z}.
    \label{C1C2}
\end{equation}
From (\ref{vmomlinear}), we find
\begin{equation}
    \partial_tv_{n} + \sum_{m=0}^\infty \left(E\partial_x v_{m}C_{1_{mn}}\right) + \beta y u_n - c_n^2 \partial_y b_n= 0
    \label{modal_eq_v},
\end{equation}
and from (\ref{buoylinear}),
\begin{equation}
    \partial_t b_{n} +\sum_{m=0}^\infty \big[ E\partial_x b_{m}C_{3_{nm}} +\beta W^2 E' v_m C_{4_{nm}}-\beta W^2 E(\partial_x u_{m}+\partial_y v_{m})C_{5_{nm}}\big] - (\partial_x u_{n} + \partial_y v_{n}) = 0,
    \label{modal_eq_b}
\end{equation}
in which
\begin{equation}
    C_{3_{nm}}=\displaystyle\frac{\displaystyle\int_{-H}^0\overline{N}^2\tilde{U}\Phi_n\Phi_m\textrm{d}z}{\displaystyle\int_{-H}^0\overline{N}^2\Phi_n^2\textrm{d}z}, \quad C_{4_{nm}}=\displaystyle\frac{\displaystyle\int_{-H}^0\overline{N}^2\tilde{U}'\Phi_n'\Phi_m\textrm{d}z}{\displaystyle\int_{-H}^0\overline{N}^2\Phi_n^2\textrm{d}z}, \quad \textrm{and} \quad C_{5_{nm}}=\displaystyle\frac{\displaystyle\int_{-H}^0\tilde{U}''\Phi_n\Phi_m\textrm{d}z}{\displaystyle\int_{-H}^0\overline{N}^2\Phi_n^2\textrm{d}z}.
    \label{C3C4C5}
\end{equation}

\section{Numerical methods}\label{num param}
The equations for the mode amplitude evolution (\ref{modal_eq_u}), (\ref{modal_eq_v}) and (\ref{modal_eq_b}) are solved using Dedalus \citep[][]{Burnsetal2020}. The coefficients (\ref{C1C2}) and (\ref{C3C4C5}) are solved beforehand, using vertical modes $\Phi_{n}$ computed with the Galerkin method for a given stratification profile from (\ref{Phieqn}), and a prescribed vertical profile $\tilde{U}$ for the jet.\\

\begin{figure}
    \centering
    \includegraphics[width=0.44\linewidth]{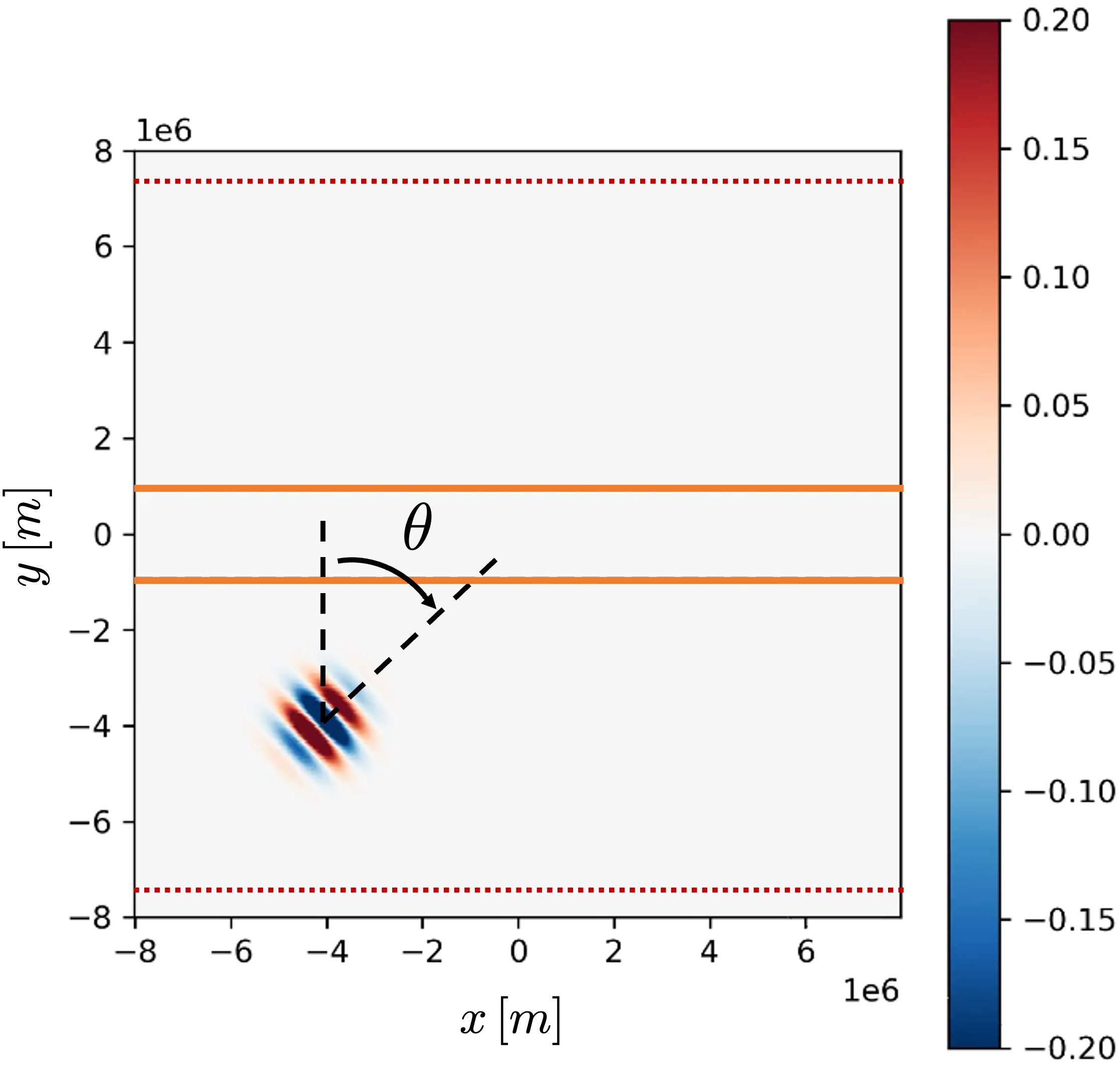}
    \captionsetup{width=1.\linewidth, justification=justified, format=plain}
    \caption{Initial condition in a typical simulation: the wavepacket is launched from the position $(x_0, y_0)$ in the southern hemisphere at an angle $\theta$ from north. Jet limits at $y\pm2$W are shown by orange lines, sponge layers limits are shown by red dashed lines.}
    \label{presentation_context2}
\end{figure}

\noindent The domain represents a patch of the Equatorial Pacific latitudinally centered on the equator. The domain is such that $x\in[-L_x;\,L_x]$ and $y\in[-L_y;\,L_y]$ where $(L_x, L_y)=(8000,8000)\,\textrm{km}$ (figure (\ref{presentation_context2})). It needs to be sufficiently large to allow the wavepacket to propagate toward the jet, interact with it and exit on the other size. The equator, specifically, is centered at $y=0$. The equatorial jet uniformly spans the zonal direction. It is centered on the equator with an horizontal width of W=400 km. The domain is periodic in both $x$ and $y$ with zonally uniform sponge layers located near the north and south boundaries of the domain to prevent the waves crossing the top and bottom boundaries.\\

\noindent The resolution is set to $(N_x, N_y)=(512,512)$ so that the wavelength of the internal tides are resolved by at least 16 grid points. The typical simulation duration is approximately 8.7 days: the number of iterations is fixed at 1500 with a timestep of $dt=500$s ensuring numerical stability of the CFL criterion. Viscosity is also added for numerical stability. It is fixed at $\nu=1$ m$^2\cdot$s$^{-1}$, though much larger than the viscosity of sea water, it is still small enough that dissipation of the waves is insignificant.\\

\noindent The dimensions of the domain are chosen to match the physics of the problem: the M2 internal tide has a forcing frequency of $\omega_0=1.4\times10^{-4}$ rad.s$^{-1}$, with typical horizontal wavelengths having an order of magnitude $\sim$400 km $(k_{h}\sim 1.5\times10^{-5}$ m$^{-1})$ and horizontal phase speeds 2 m$\cdot$s$^{-1}$. We are free to choose the orientation $\theta$ from north of initial horizontal wavenumber vector $(k_{x}, k_{y})=(k_{h}\textrm{sin}\theta, k_{h}\textrm{cos}\theta)$.\\

\noindent The model is tested here for a constant buoyancy frequency of $\overline{N}_{0}^2=1\times10^{-4}$ s$^{-2}$. Through the dispersion relation (\ref{relation_dispersion_toymodel}), for a mode 1 wave at $y_{0}\ll-$W the resulting M2 horizontal wavelength is of order 600 km, a value on the larger side.\\

\noindent The wavepacket is initialized with inner oscillations of $\omega_0$ fixed at the M2 forcing frequency (figure (\ref{presentation_context2})). The effect of the rotation is included using $f=\beta y_{0}$. The envelope has a typical width of $\sigma_x$ and $\sigma_y$ with an amplitude fixed as $A_0=0.01\times H$ with $H$ the depth of the ocean previously used for computing the vertical modes. The wavepacket can be launched from different locations $(x_0, y_0)$ and with various angles $\theta$ compared to the north.\\

\noindent The kinetic energy (KE), available potential energy (APE) and total energy (TE=KE+APE) of the different modes of the internal tides are computed numerically using the vertical mode decomposition:
\begin{equation}
    \textrm{KE}_n=\frac{1}{2}\!\int\!\int(u_n^2+v_n^2)\textrm{d}x\textrm{d}y\int(\Phi{_n}^2)'\textrm{d}z,
    \label{num_KE}
\end{equation}
\begin{equation}
    \textrm{APE}_n=\frac{1}{2\overline{N}_0^2}\!\int\!\int b_n^2\textrm{d}x\textrm{d}y\int\overline{N}_0^2\Phi{_n}^2\textrm{d}z.
    \label{num_APE}
\end{equation}

\noindent The total energy budget is also computed:
\begin{equation}
	\partial_t(KE+APE)=-uv\partial_y U-uw\partial_zU-\frac{1}{\overline{N}^2}bv\partial_yB-\frac{1}{\overline{N}^2}bw\partial_zB,
	\label{bilan_energie}
\end{equation}
where the two first terms of the right hand side correspond to shear production and the two others to buoyancy production.\\

\section{Results for the model $\rm{\overline{N}}_{0}=$ constant} \label{toymodel}
\subsection{No Equatorial jet}
\subsubsection{Trajectories and dynamics}
\begin{figure}
    \centering
    \includegraphics[width=0.48\linewidth]{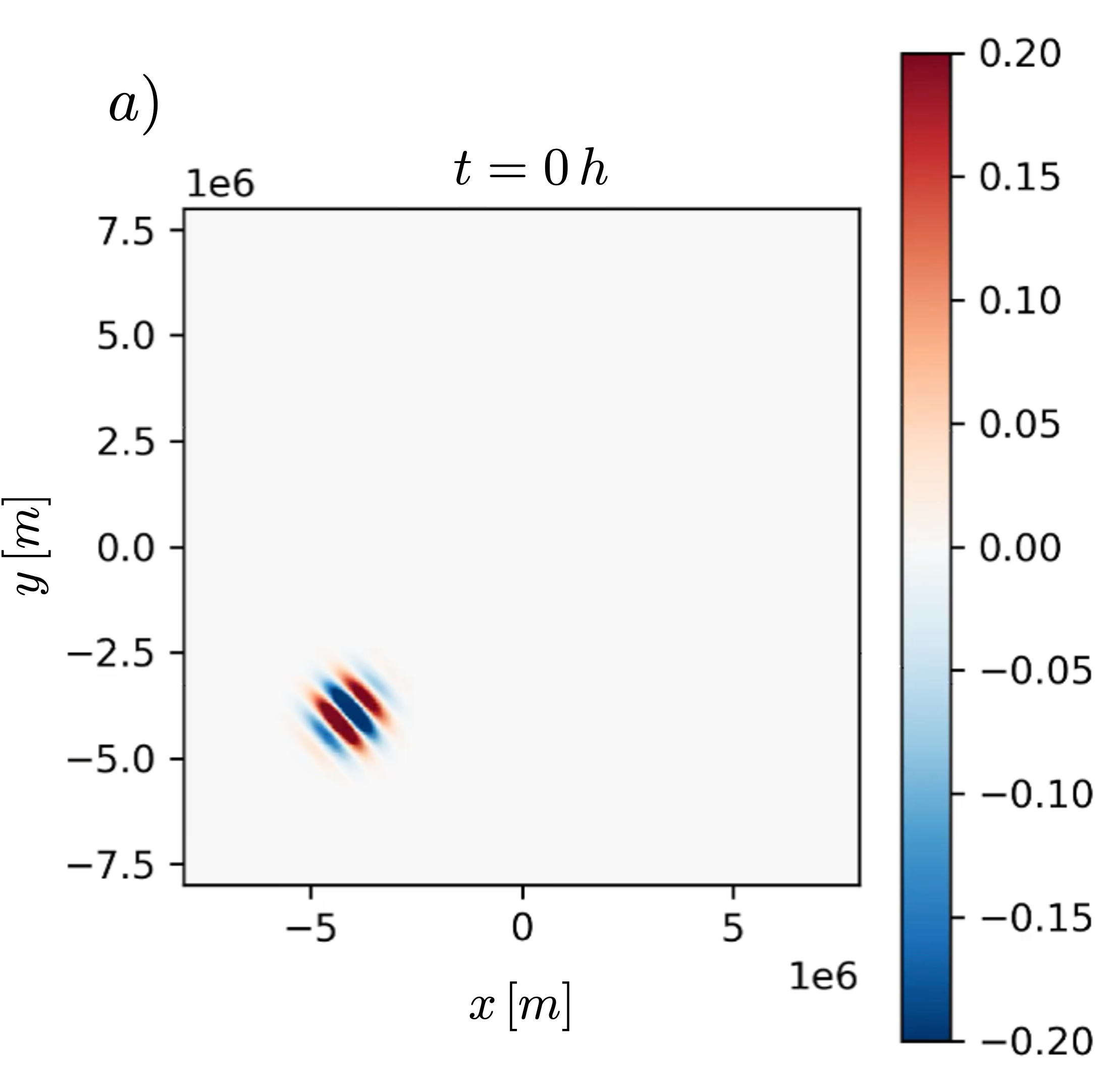}
    \includegraphics[width=0.465\linewidth]{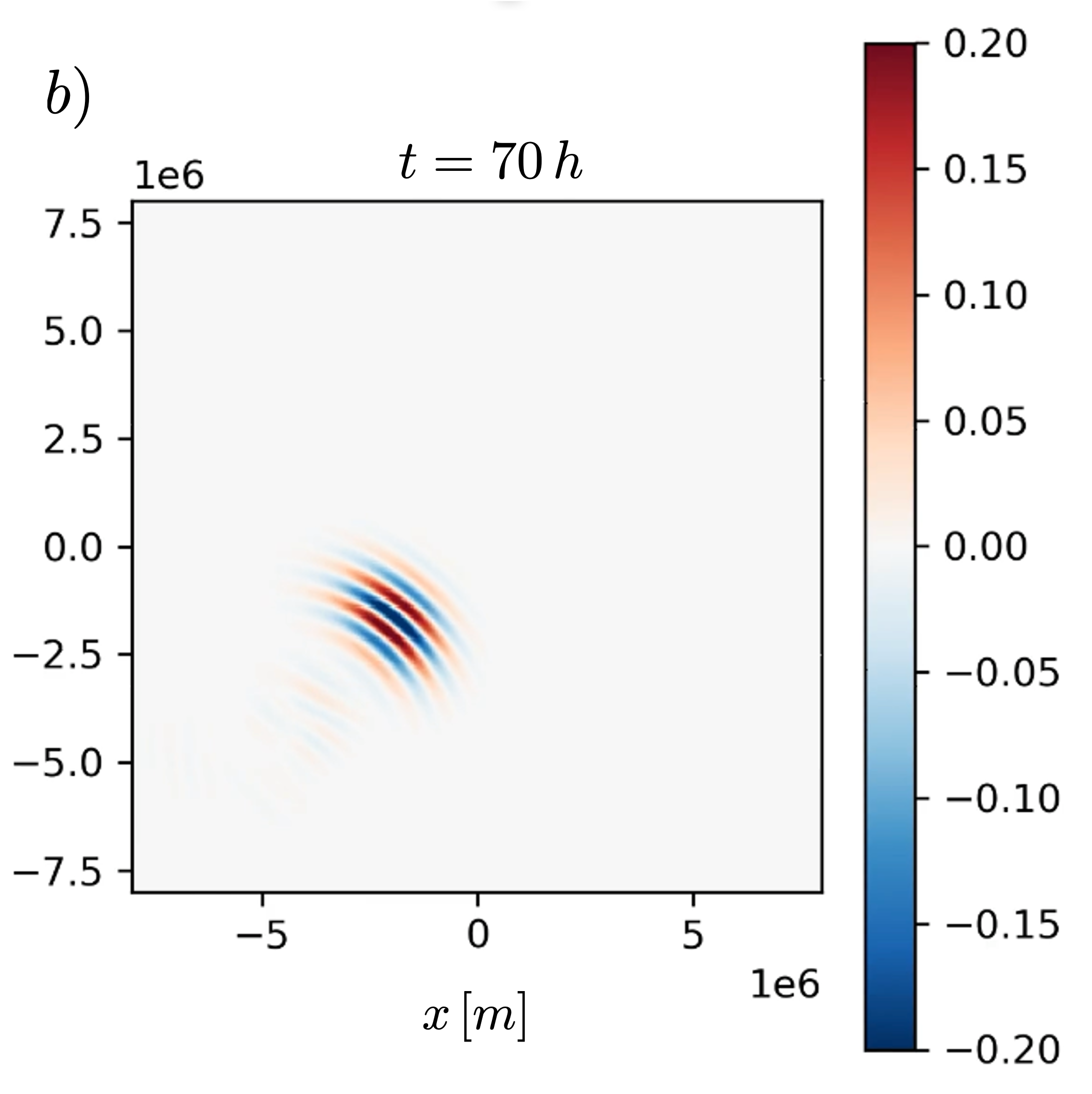}
    \includegraphics[width=0.48\linewidth]{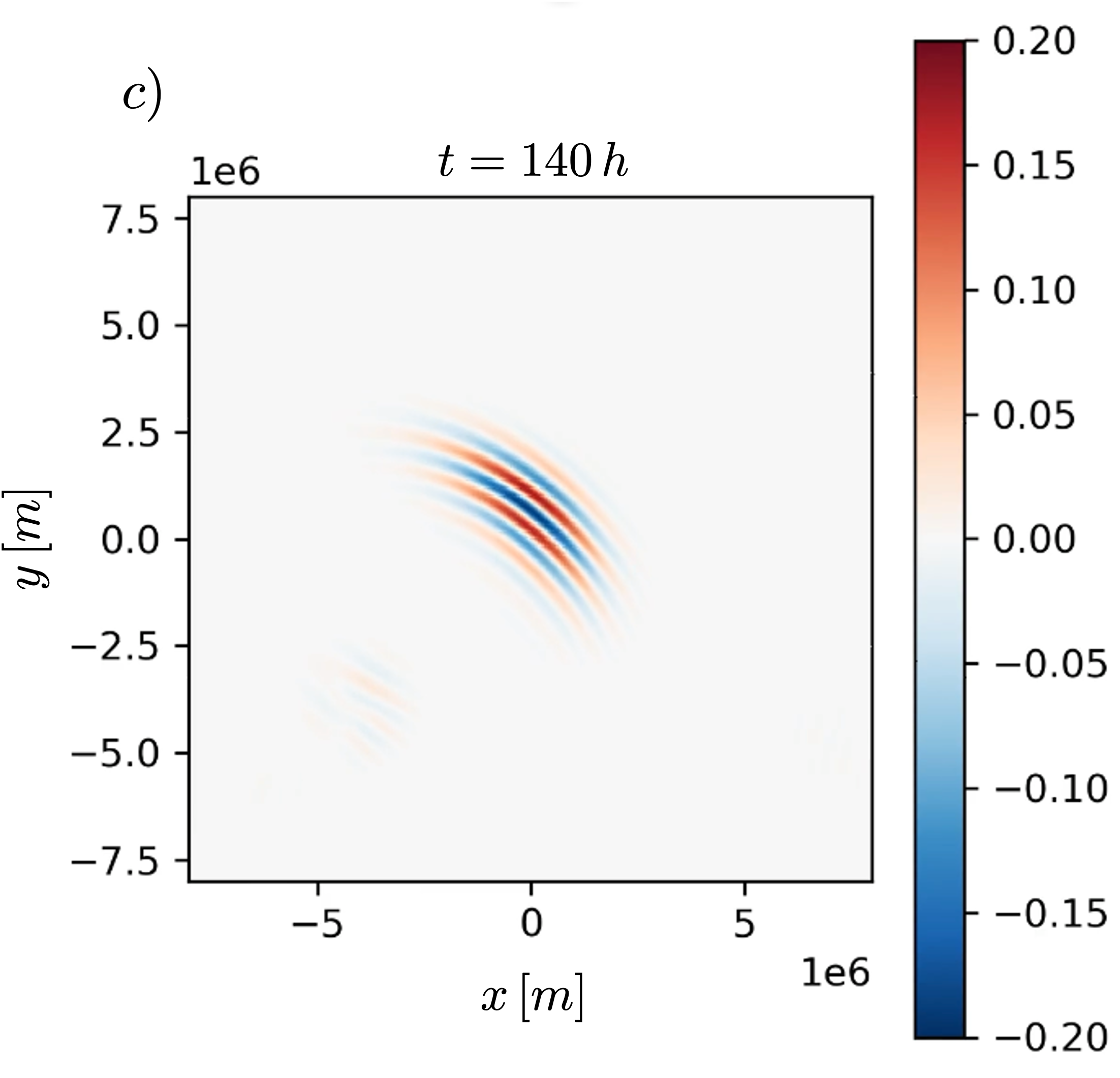}
    \includegraphics[width=0.465\linewidth]{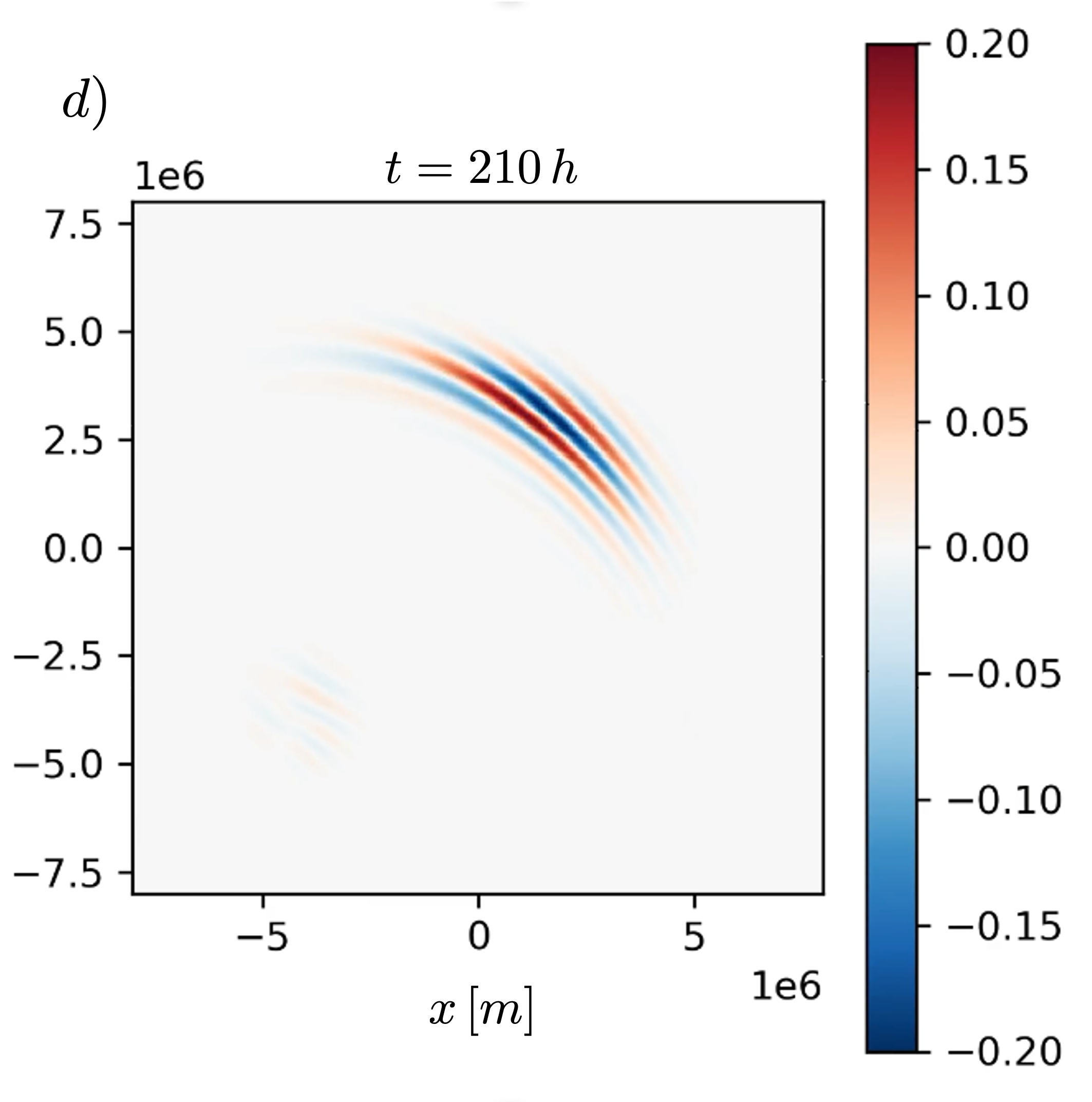}
    \captionsetup{width=1.\linewidth, justification=justified, format=plain}
    \caption{Zonal velocity (m$\cdot$s$^{-1}$) of vertical mode 1 M2 internal tide. The wavepacket is launched from $(x_0, y_0)=(-0.5\,L_x, -0.5\,L_y)$ with $\theta=\pi/4$.}
    \label{test_dynamic_nojet}
\end{figure}
The model is first tested without any equatorial jet, to check the propagation of the wavepacket and the influence of the rotation on its dynamics through the beta-plane approximation. In all that follows, the buoyancy frequency is kept constant, at $\overline{N_0}^2=1\times10^{-4}$ s$^{-2}$. In that case, (\ref{modal_eq_u}), (\ref{modal_eq_v}) and (\ref{modal_eq_b}) reduce to decoupled equations for each mode:
\begin{subequations}
\begin{align}
    \partial_tu_{n} - \beta y v_n - c_n^2 \partial_x b_n =0, \\
    \partial_tv_{n} + \beta y u_n - c_n^2 \partial_y b_n =0, \\
    \partial_tb_{n} - ( \partial_xu_n + \partial_yv_n) =0,
\end{align}
\label{system_no_jet}
\end{subequations}
where $c_n^2=(\omega_0^2-(\beta y)^2)/k_h^2$.\\

\noindent For a mode 1 wave $(n=1)$, $c_{1}$ is fixed. Setting $\omega_{0}=1.4\times 10^{-4}$ rad.s$^{-1}$  yields the initial wavenumber $k_{h0}=7.2\times10^{-6}$ m$^{-1}$.\\

\noindent Figure (\ref{test_dynamic_nojet}) shows the evolution of a wavepacket, launched from $(x_0, y_0)=(-Lx/2, -Ly/2)=(-4000,-4000)$km with an angle of $\theta=\pi/4$ and $\sigma_{x}=\sigma_{y}=500$ km. As such, the wavepacket is quasi-monochromatic, with $\sigma k_{h0}\gg 1$. As it crosses the domain, its trajectory slightly deviated. It is also subjected to significant dispersion due to the limited number of wavelengths composing it. The measurements over time in simulations yield values around $c_{p}=27$m$\cdot$s$^{-1}$ and $c_{g}=9$m$\cdot$s$^{-1}$ for the mode 1.

\subsubsection{Ray tracing}
The ray tracing analysis (see section \ref{ray_tracing_general} with $\omega_{\textrm{int}}=\omega_{0}$) can help disentangle the effect of the rotation over the wavepacket. In the $\beta$-plane approximation, the trajectory of the wavepacket can be computed using (\ref{ray_tracing_traj_formula}) with $E(y)=0$:
\begin{equation}
    x(y)=\frac{1}{\sqrt{C_1\beta^2}}\,\textrm{atan}\left(\frac{y}{\sqrt{C_2-y^2}}\right),
    \label{traj_raytracing_nojet}
\end{equation}
with:
\begin{equation}
    C_1=\frac{\pi^2}{H^2k_{x_0}^2(\overline{N_0}^2-\omega_0^2)}, \quad \textrm{and} \quad C_2=\frac{1}{\beta^2}\left(\omega_0^2-\frac{1}{C_1}\right)
\end{equation}
\noindent This expression is valid provided:
\begin{equation}
    y^2<C_{2}=\frac{\omega_0^2}{\beta^2}-\frac{1}{C_1\beta^2}
\end{equation}

\begin{figure}
  \centering
   \includegraphics[width=0.48\linewidth,trim=0cm 0cm 0cm 0.2cm,clip]{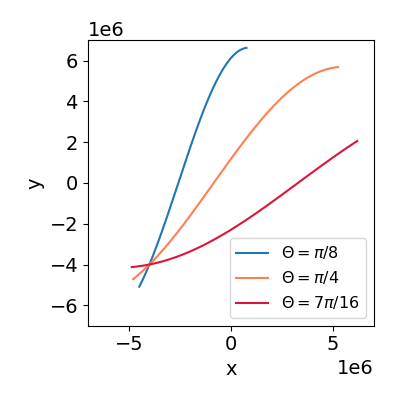}
   \caption{Trajectories computed using ray tracing for different angles of launch $\theta$.}
  \label{trajectoire_raytracing_nojet}
\end{figure}
\noindent The effect of the rotation and the influence of the initial angle of launch both directly influence the trajectory of the wavepacket through (\ref{traj_raytracing_nojet}).\\
Figure (\ref{trajectoire_raytracing_nojet}) shows the expected trajectories for angles of $\theta=\pi/4$, $\theta=\pi/8$ and $\theta=7\pi/16$. Rotation is responsible for the observed deviation of the trajectories. This effect also tends to disappear as the wavepacket approaches and crosses the Equator (where $\beta y$ goes to zero).\\

\begin{figure}
    \centering
    \includegraphics[width=0.48\linewidth]{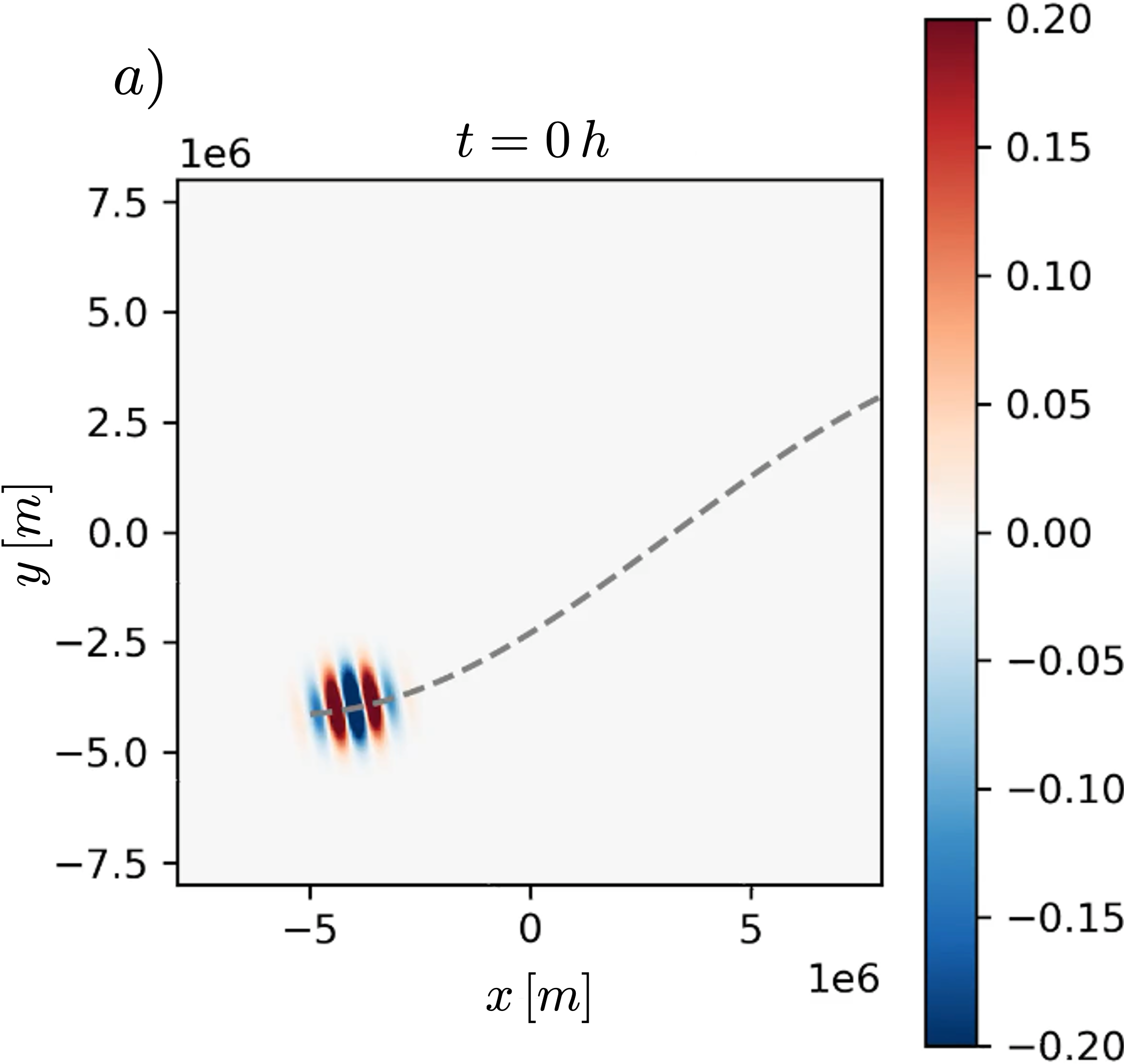}
    \includegraphics[width=0.47\linewidth]{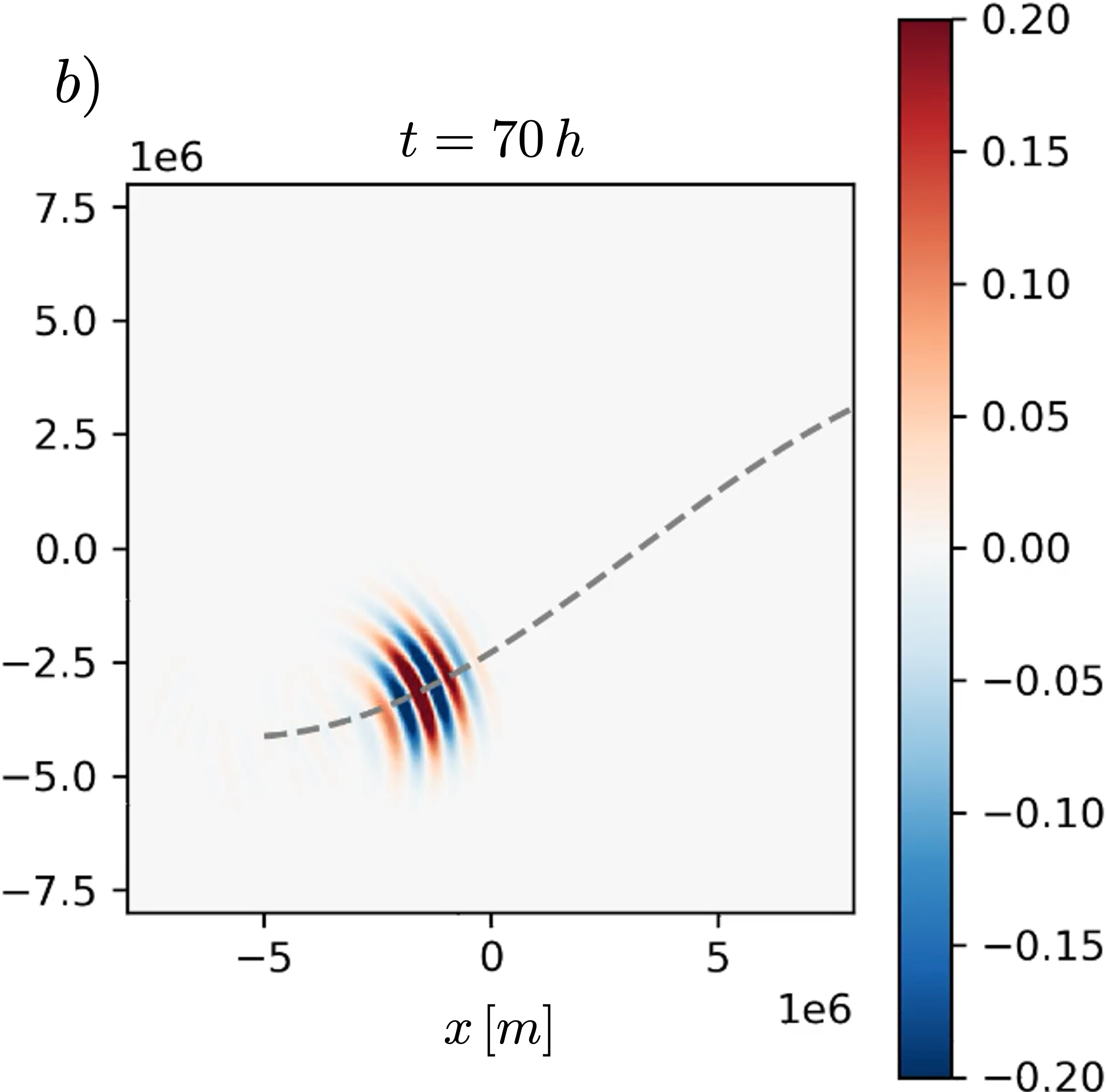}
    \includegraphics[width=0.489\linewidth]{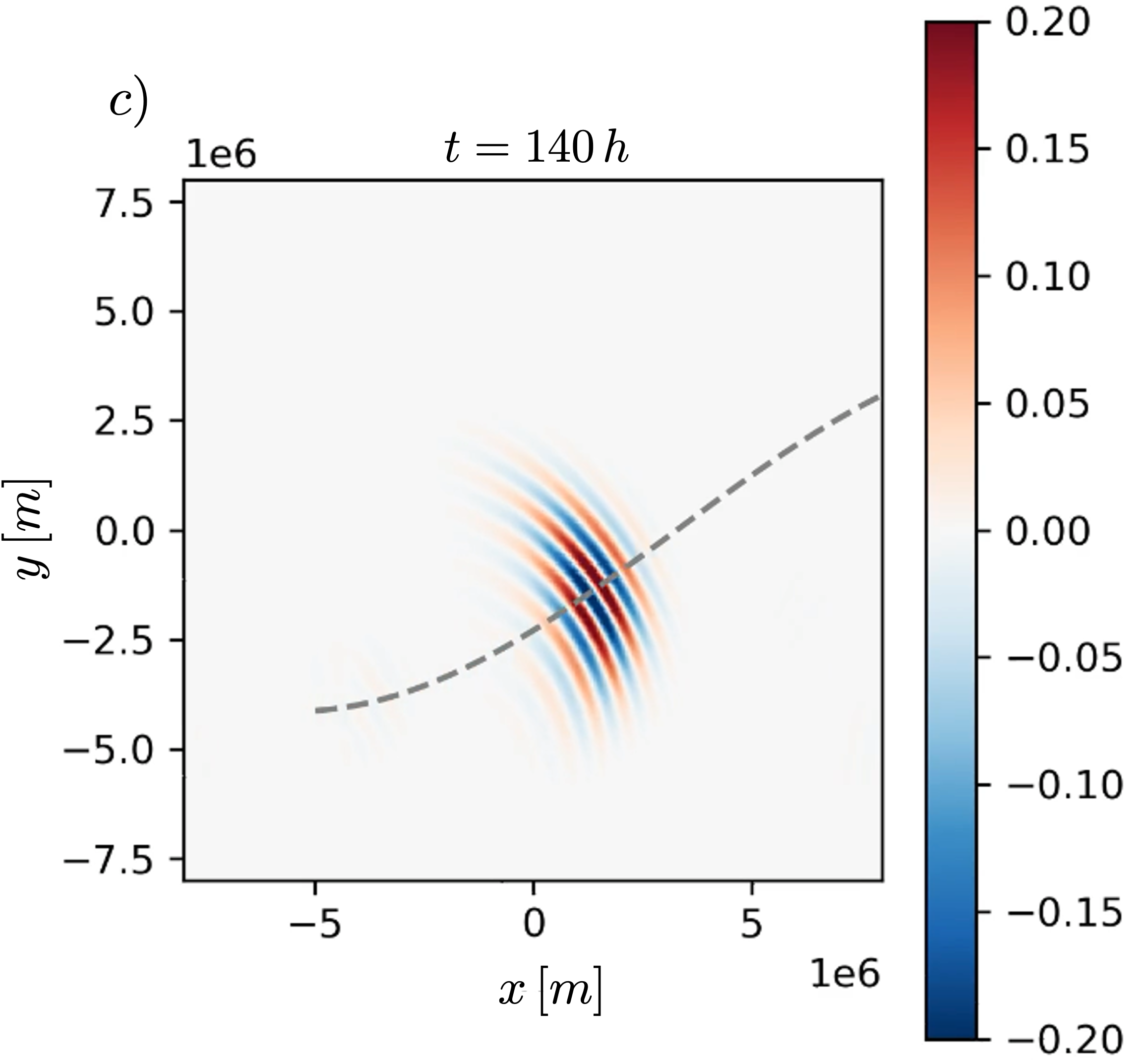}
    \includegraphics[width=0.465\linewidth]{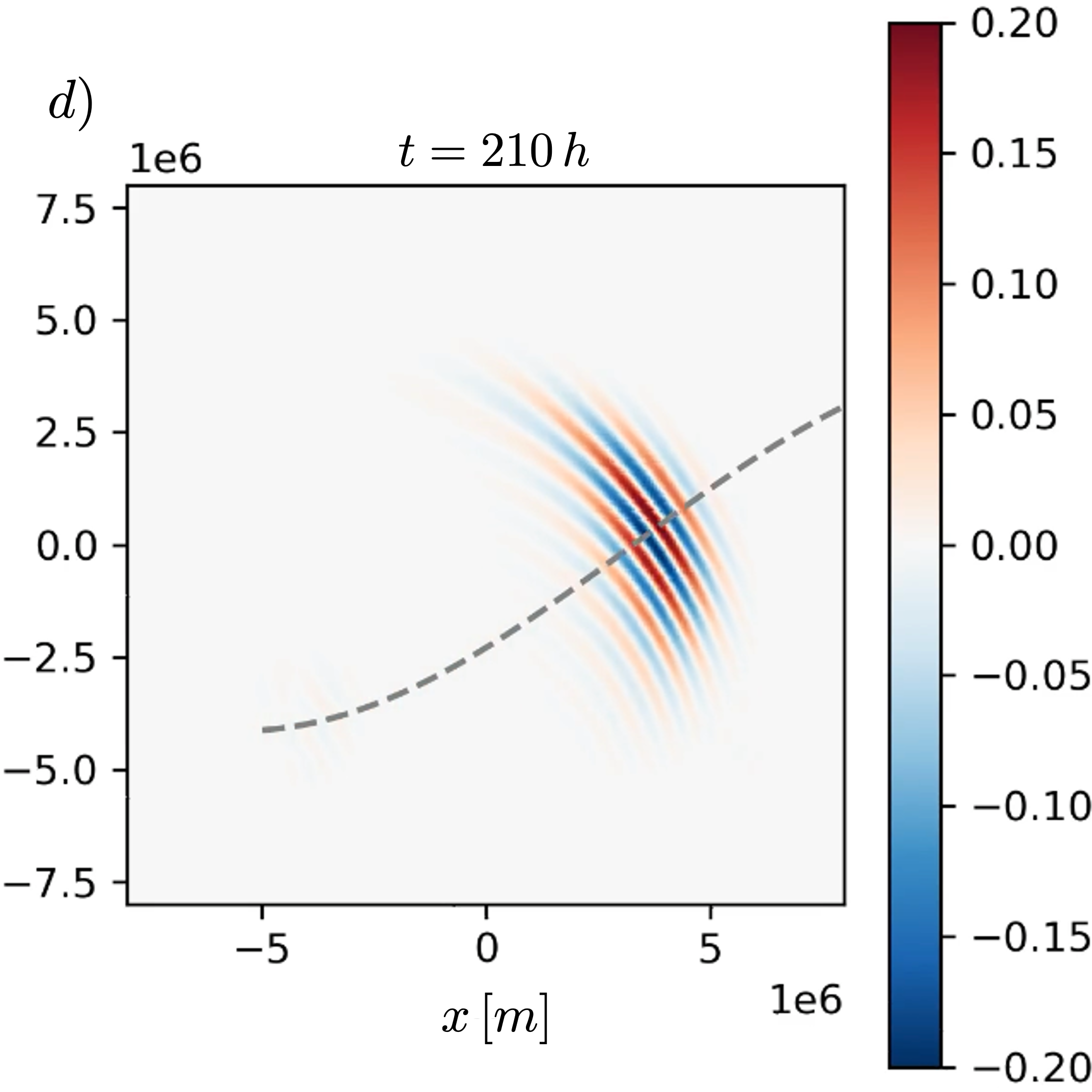}
    \captionsetup{width=1.\linewidth, justification=justified, format=plain}
    \caption{Colors: zonal velocity (m$\cdot$s$^{-1}$) of vertical mode 1 M2 internal tide. The wavepacket is launched from $(x_0, y_0)=(-0.5\,L_x, -0.5\,L_y)$ with $\theta=7\pi/16$. Grey dashed lines: trajectory computed using ray tracing.}
    \label{traj_raytracing_simulation_nojet}
\end{figure}

\noindent Figure (\ref{traj_raytracing_simulation_nojet}) shows the superposition of the computed trajectory with an initial angle of $\theta=7\pi/16$ and the simulation. Both are in good agreement and illustrate how ray tracing succeeds in predicting the path of the wavepacket due to the rotation.
\subsubsection{Energetics}
In addition to the numerical computation (\ref{num_KE}), KE was computed analytically:
\begin{equation}
    \textrm{KE}_{1}=\frac{\pi}{16}\textrm{A}_0^2\sigma_x\sigma_y\left(\overline{N}^2+2\,\frac{m_{1}^2}{k_h^2}(\beta y)^2\right),
    \label{KE_analytics}
\end{equation}
Equation (\ref{KE_analytics}) shows how rotation directly acts as a source term of kinetic energy for internal tides.\\

\begin{figure}
    \centering
    \includegraphics[width=\linewidth]{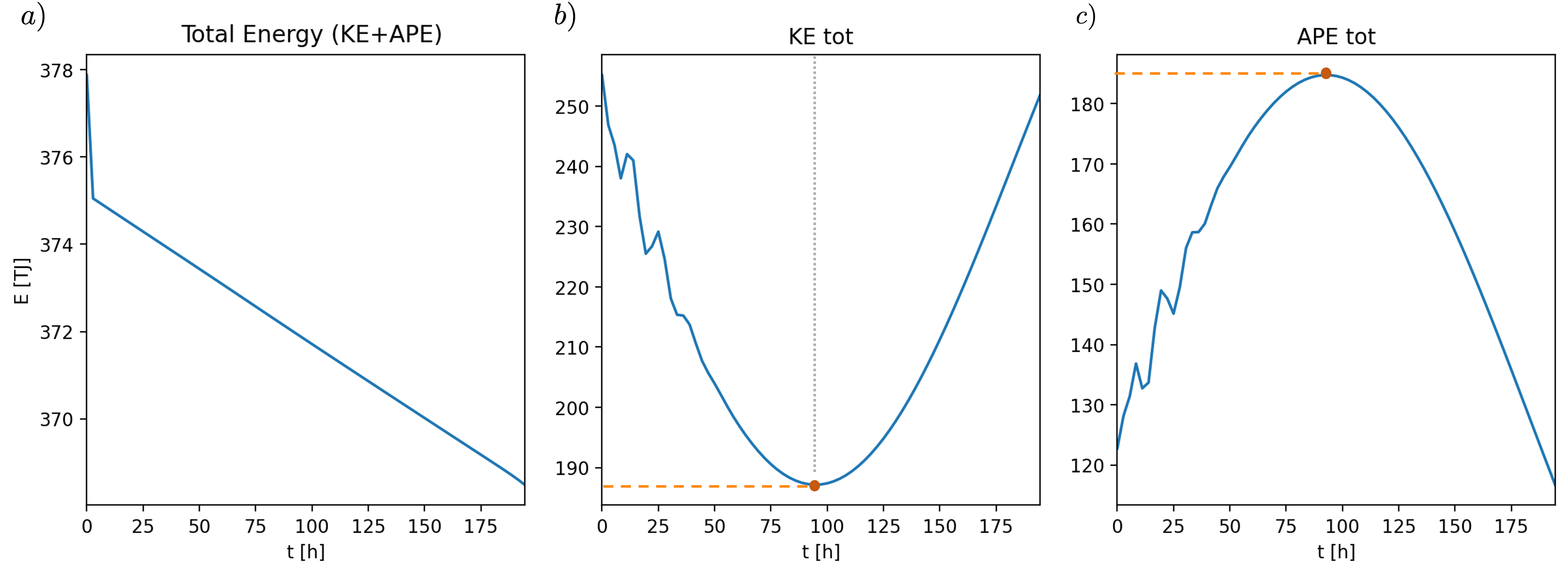}
    \captionsetup{width=1.\linewidth, justification=justified, format=plain}
    \caption{a) total energy, b) kinetic energy, c) available potential energy; all computed for the simulation presented in figure (\ref{test_dynamic_nojet}).}
    \label{energetics_nojet}
\end{figure}

\noindent Figure (\ref{energetics_nojet}) shows the total energy (TE; left panel), kinetic energy (KE; middle panel) and available potential energy (APE; right panel) of the first mode of the internal tide corresponding to the case in figure (\ref{test_dynamic_nojet}). As only the first mode is initialized and because there is no jet to create new modes, all the wave energy is present in the first mode.
The steady decrease of total energy (figure (\ref{energetics_nojet}), left panel) is explained by the dissipation added to guarantee numerical stability (fixed at $\nu=1$ m$\cdot$s$^{-2}$). Over the time of the simulation, the loss of energy due to viscosity is $\sim$1\% initial energy.
The kinetic energy (figure (\ref{energetics_nojet}), middle panel) first decreases as the wavepacket approaches the Equator as a result of the term due to rotation in (\ref{KE_analytics}) tending to zero. Once the tide has crossed the Equator, the effect of rotation increases and KE increases again.
APE (figure (\ref{energetics_nojet}), right panel) changes as KE changes to keep the total energy nearly constant. In particular, at the Equator (here at $t\sim90$h), equipartition of the energy between KE and APE is reached.

\subsection{Vertically uniform equatorial jet}
\subsubsection{Jet configuration and M2 dynamics}
The equatorial jet is added, centered on $y=0$ with a characteristic width of $\textrm{W}=400$ km and is uniform in the vertical:
\begin{equation}
    U=U_{0}\textrm{exp}\left(-\frac{y^2}{2\rm{W}^2}\right).
    \label{uniform_jet_structure}
\end{equation}
The strength of the jet and its direction is fixed by the value taken for U$_0$ (positive eastward and negative westward).\\

\noindent Following (\ref{uniform_jet_structure}), the equations (\ref{modal_eq_u}), (\ref{modal_eq_v}) and (\ref{modal_eq_b}) become:
\begin{subequations}
\begin{align}
    \partial_tu_{n}+EU_0\partial_x u_n+v_n E'U_0- \beta y v_n - c_n^2 \partial_x b_n =0,\label{system_uniform_jet_a} \\
    \partial_tv_{n}+EU_0\partial_x v_n + \beta y u_n - c_n^2 \partial_y b_n =0, \\
    \partial_tb_{n}+EU_0\partial_x b_n - ( \partial_xu_n + \partial_yv_n) =0.
\end{align}
\label{system_uniform_jet}
\end{subequations}
Again the lack of vertical structure to $U$ decouples the mode-mode interactions. The added terms show that the flow mostly contributes to the advection of the wave. Only the term $v_n E'U_0$ in (\ref{system_uniform_jet_a}) contributes to the interaction between the perturbations and the mean flow gradient.\\

\subsubsection{Ray tracing}
The trajectories of the wavepacket can be predicted with ray tracing (see section \ref{ray_tracing_general}, with $\omega_{0}=\omega_{\textrm{int}}+Uk_{x}$). As the jet is located at the equator, where the rotation through the $\beta$-plane approximation is negligible, the Coriolis force is not included; $\omega_{0}=\omega_{\textrm{int}}(k_{x}, k_{y})+Uk_{x}$ with $\omega_{\textrm{int}}(k_{x}, k_{y})$ the dispersion relation (\ref{relation_dispersion_toymodel}), which is then independent of $y$. Using (\ref{ray_tracing_traj_formula}), we find 
\begin{equation}
    x(y)=k_{x_0}\left(\frac{c_{gx_{\textrm{int}}}}{c_{gy_{\textrm{int}}}}y+\frac{1}{c_{gy_{\textrm{int}}}}\sqrt{\frac{\pi}{2}}\textrm{W}\,\textrm{erf}\left(\frac{y}{\sqrt{2}\textrm{W}}\right)\right)
\end{equation}

\begin{figure}
    \centering
    \includegraphics[width=0.489\linewidth]{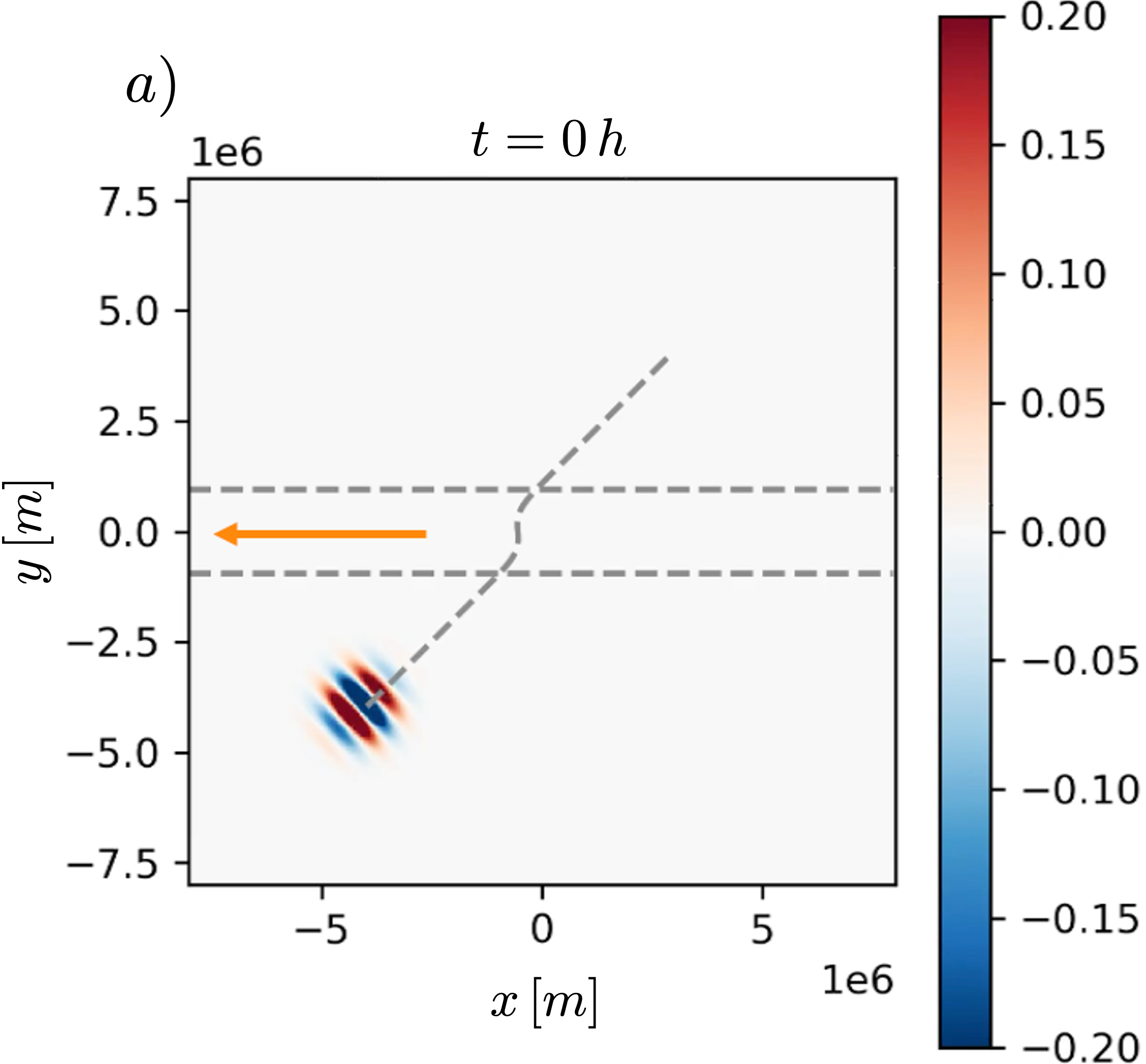}
    \includegraphics[width=0.465\linewidth]{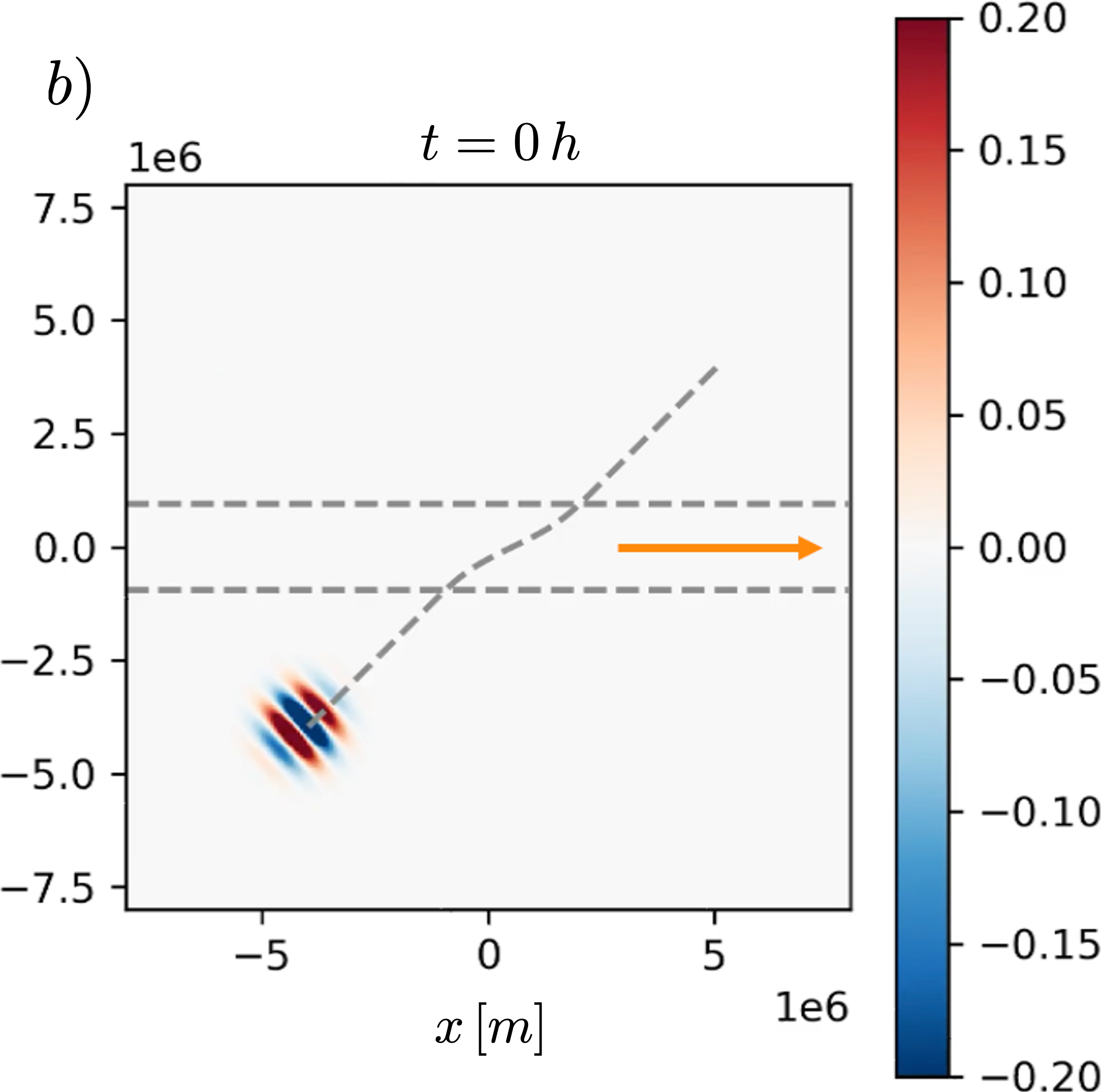}
    \captionsetup{width=1.\linewidth, justification=justified, format=plain}
    \caption{Zonal velocity (m$\cdot$s$^{-1}$) of vertical mode 1 M2 internal tide. The wavepacket is launched from $(x_0, y_0)=(-0.5\,L_x, -0.5\,L_y)$ with $\theta=\pi/4$. Trajectory computed using ray tracing shown by oblique grey dashed lines. Jet limits are denoted by horizontal grey dashed lines. Jet is uniform on the vertical with maximum amplitude set at U$_0=5$m$\cdot$s$^{-1}$, its direction varies according to the orange arrow: a) westward and b) eastward.}
    \label{trajectories_uniformjet}
\end{figure}
\noindent The trajectories for two jets of opposite direction ($U_0$=-5 m$\cdot$s$^{-1}$ and $U_0$=5m$\cdot$s$^{-1}$) are shown in figure (\ref{trajectories_uniformjet}). These results are valid only inside the jet, where the effect of the rotation is negligible. Depending on the direction of the jet and its intensity, the trajectory of the wavepacket is diverted in the direction opposite (case $U_0$=-5) or toward $(U_{0}=5)$ the direction propagation.\\

\begin{figure}
    \centering
    \includegraphics[width=0.489\linewidth]{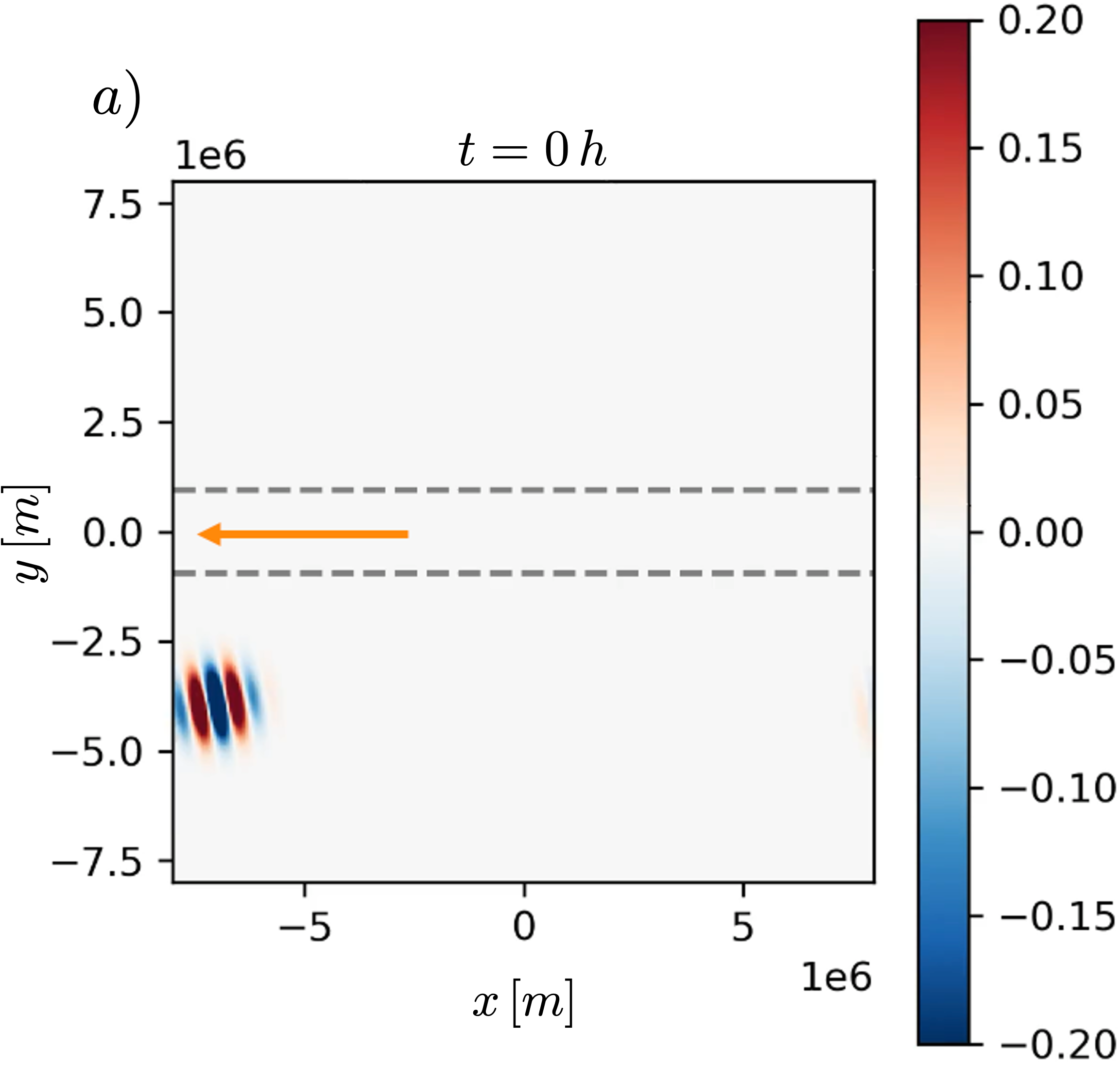}
    \includegraphics[width=0.465\linewidth]{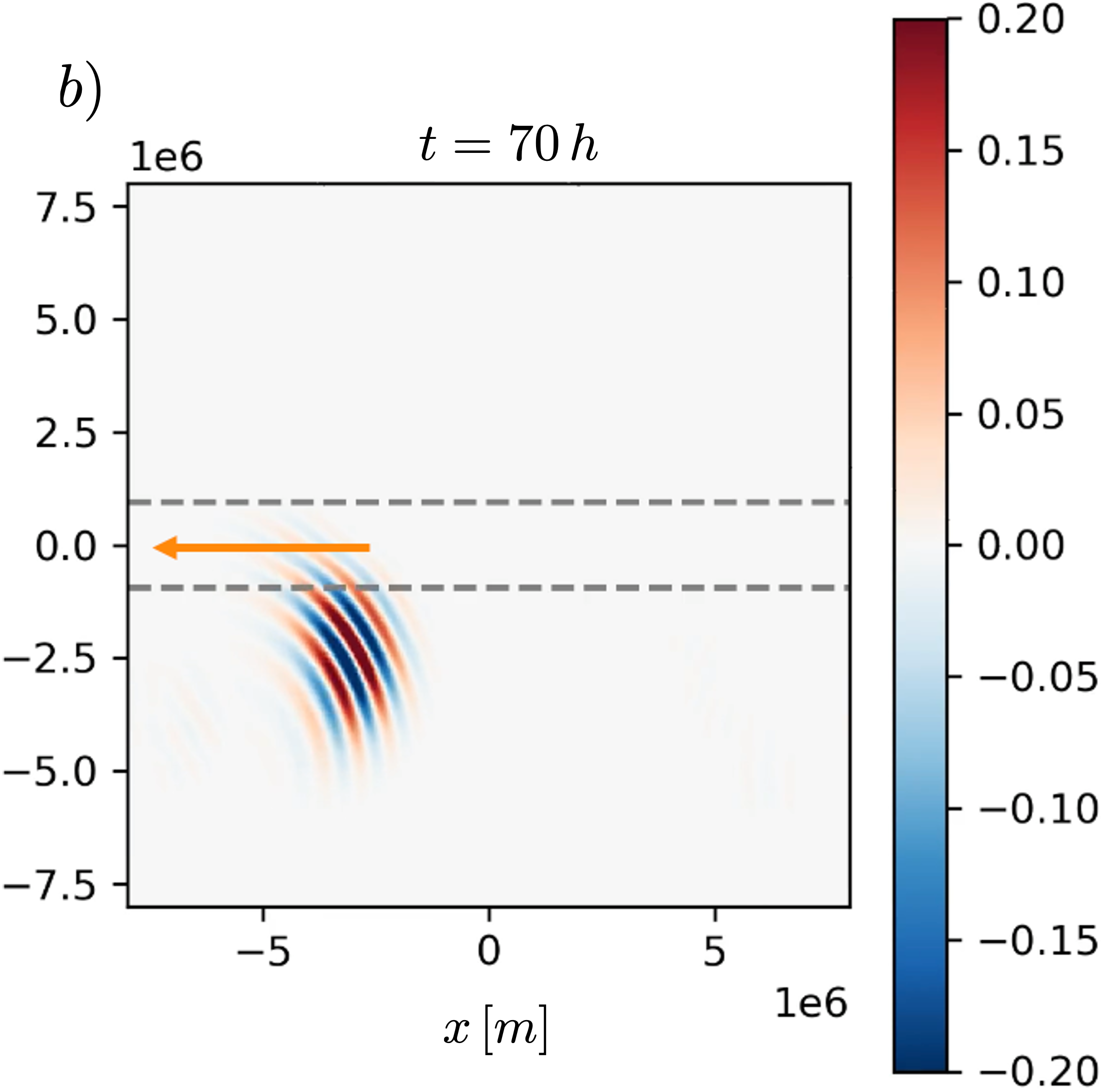}
    \includegraphics[width=0.489\linewidth]{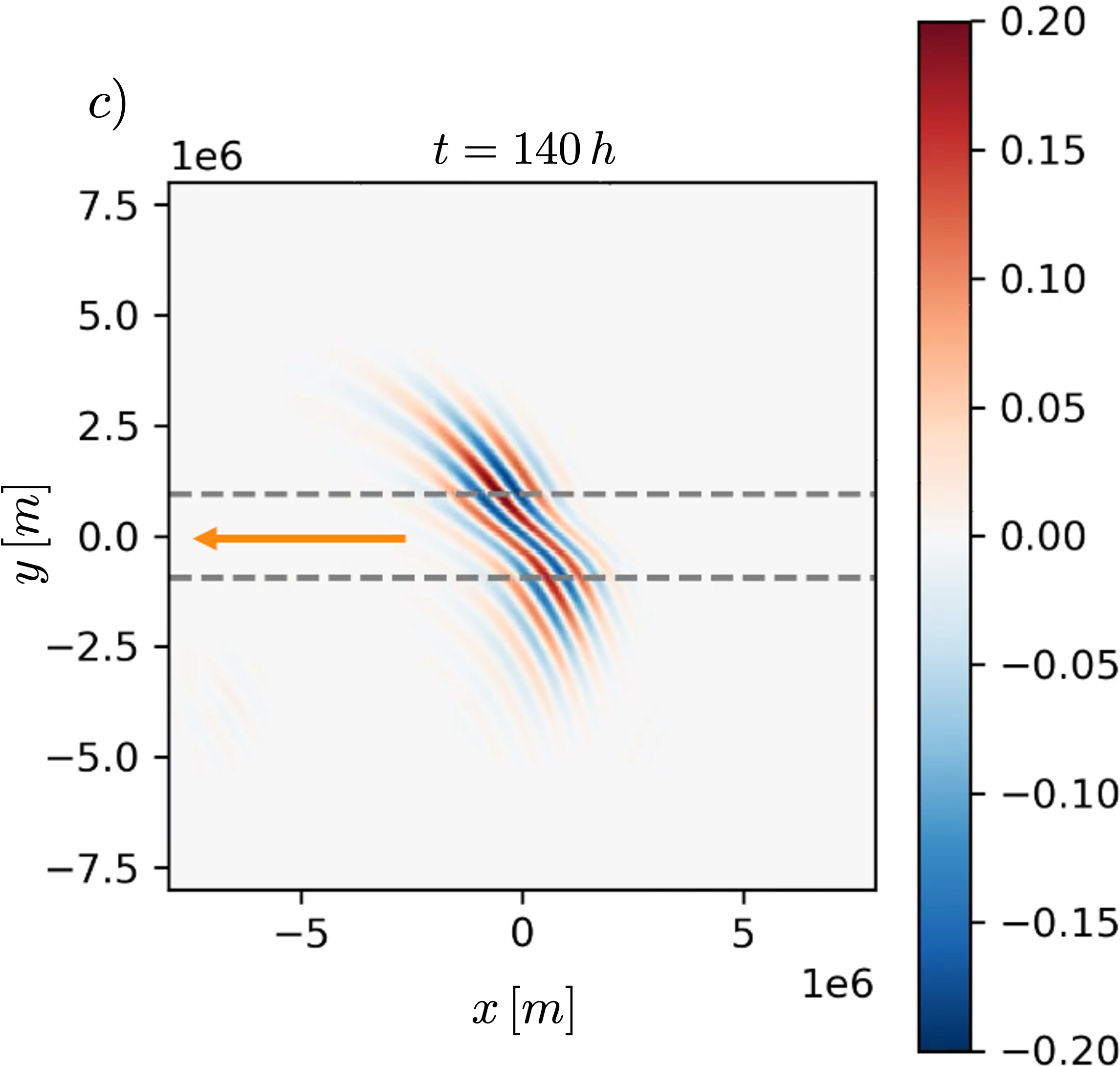}
    \includegraphics[width=0.465\linewidth]{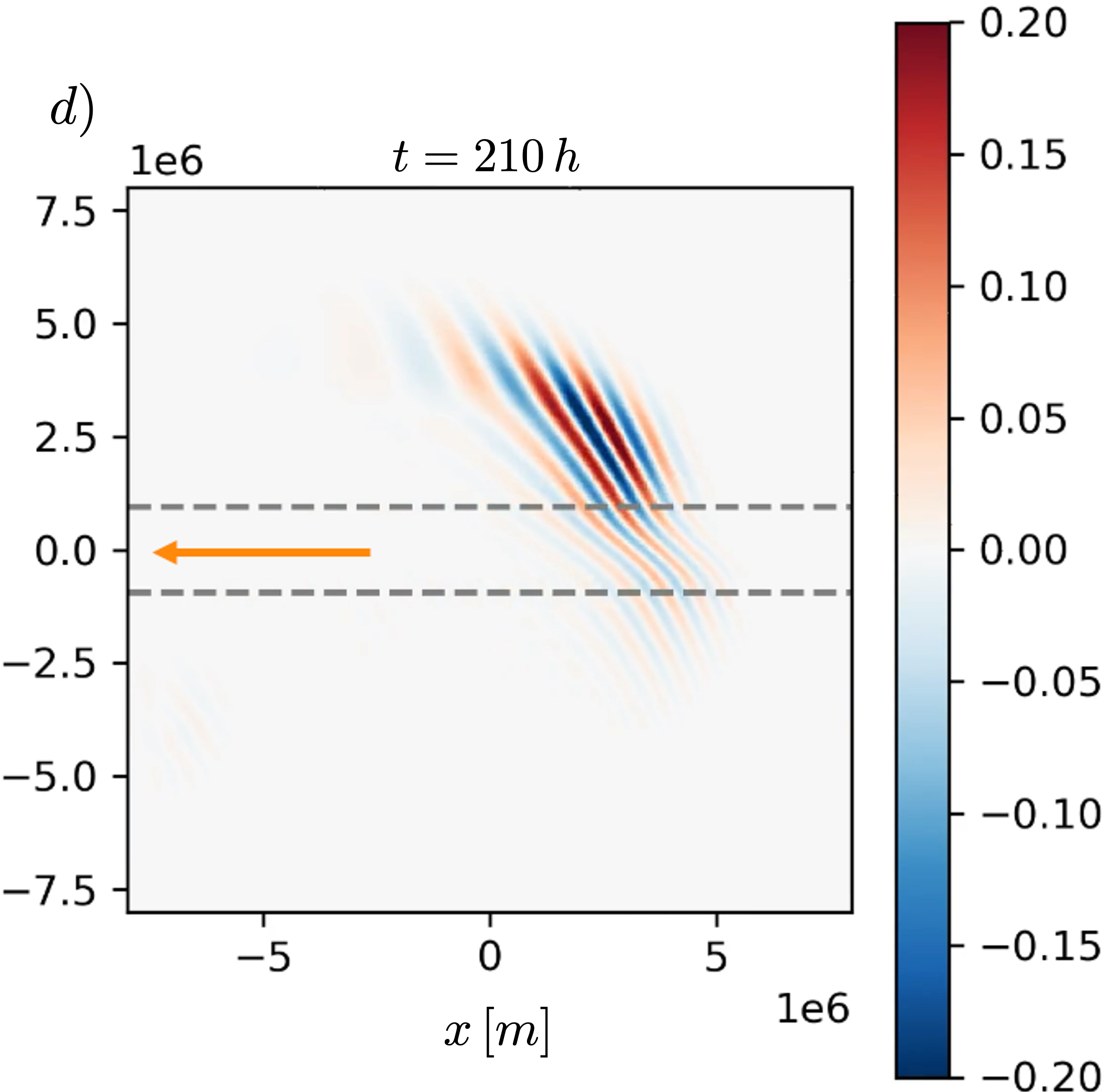}
    \captionsetup{width=1.\linewidth, justification=justified, format=plain}
    \caption{Zonal velocity (m$\cdot$s$^{-1}$) of vertical mode 1 M2 internal tide. The wavepacket is launched from $(x_0, y_0)=(-0.9\,L_x, -0.5\,L_y)$ with $\theta=7\pi/16$. Jet limits are denoted by horizontal grey dashed lines, its direction by the orange arrow. Jet is uniform on the vertical with U$_0=-6$m$\cdot$s$^{-1}$.}
    \label{critical_layer_uniformjet}
\end{figure}
\begin{figure}
    \centering
    \includegraphics[width=0.489\linewidth]{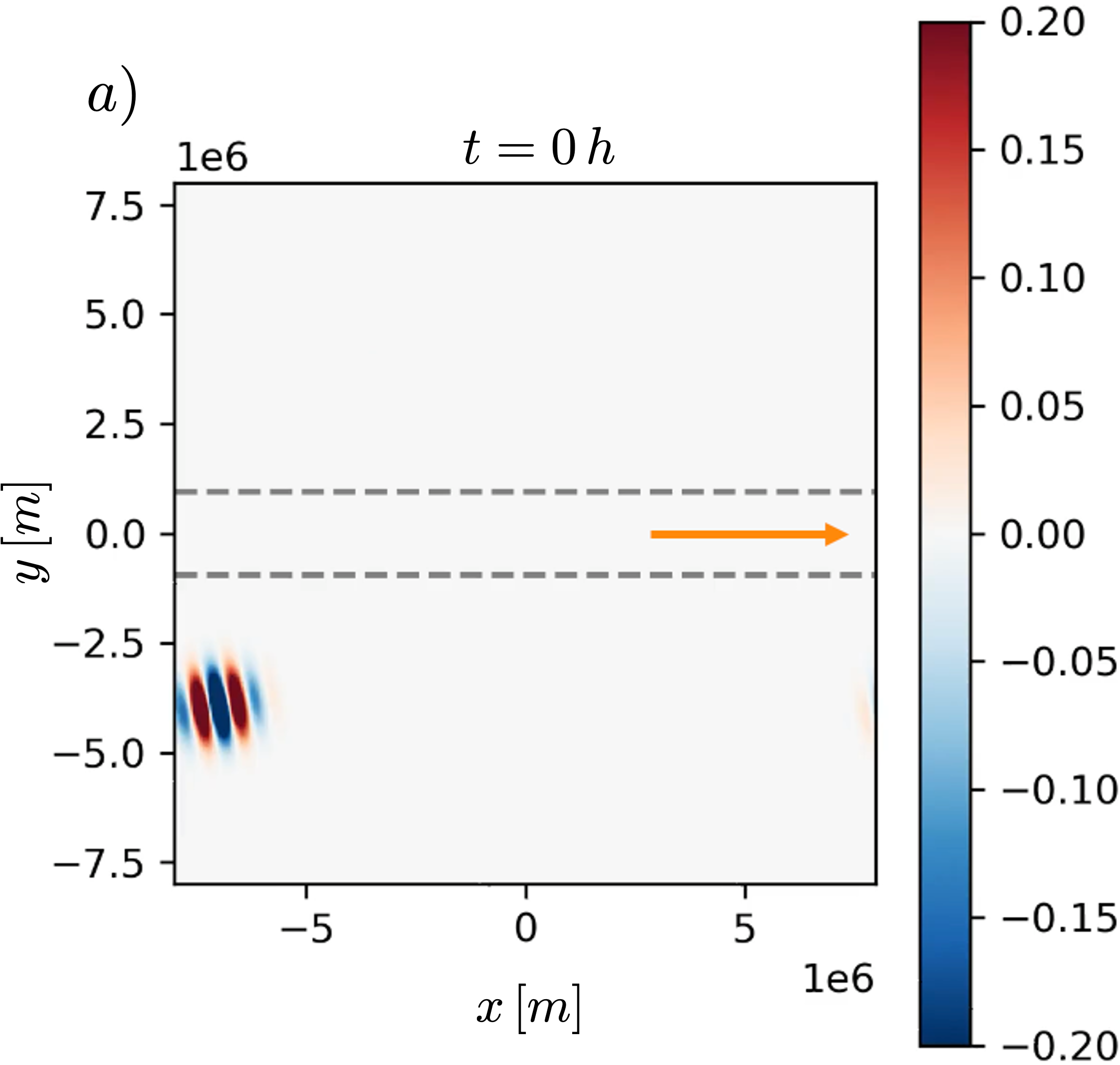}
    \includegraphics[width=0.465\linewidth]{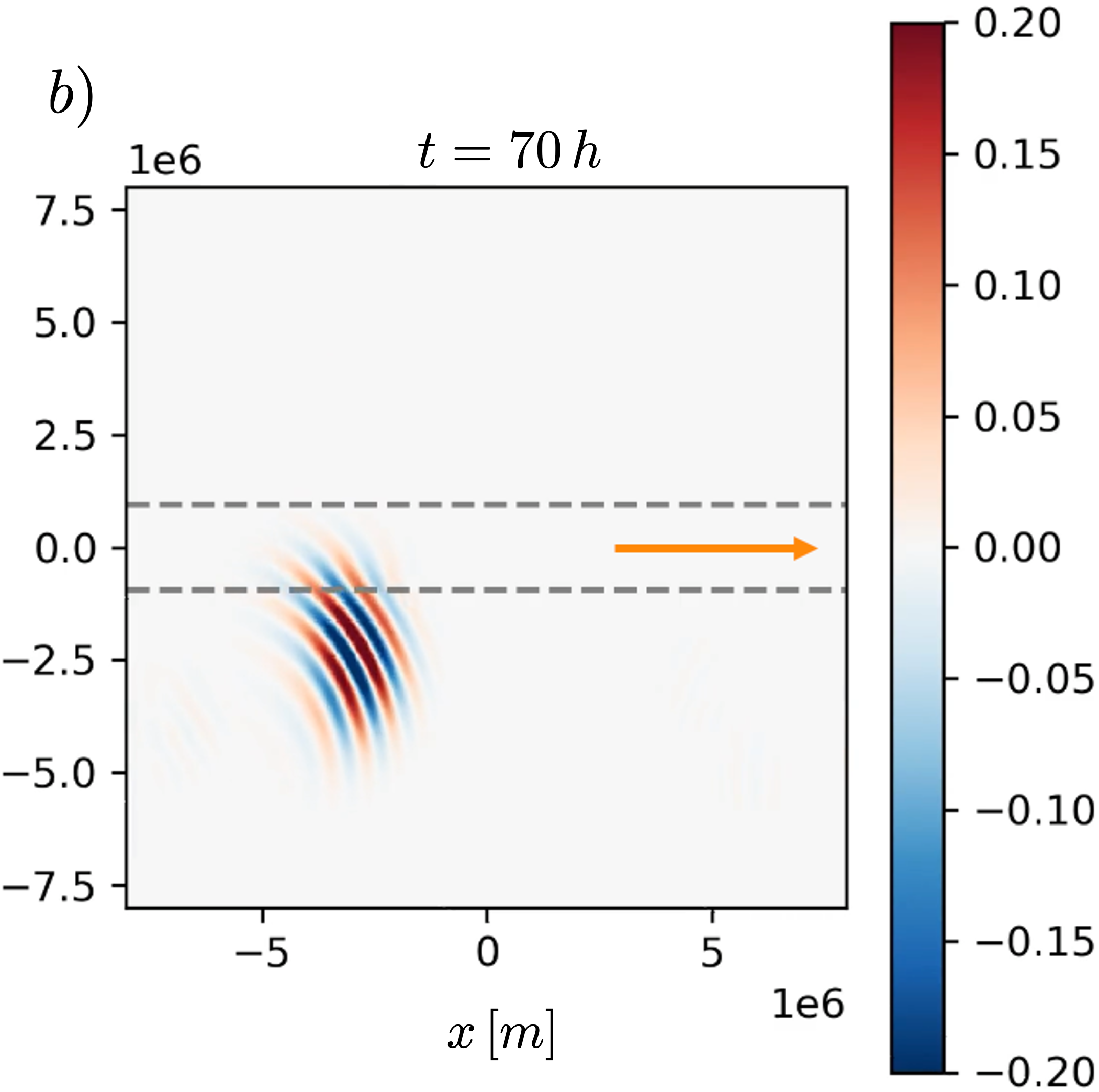}
    \includegraphics[width=0.489\linewidth]{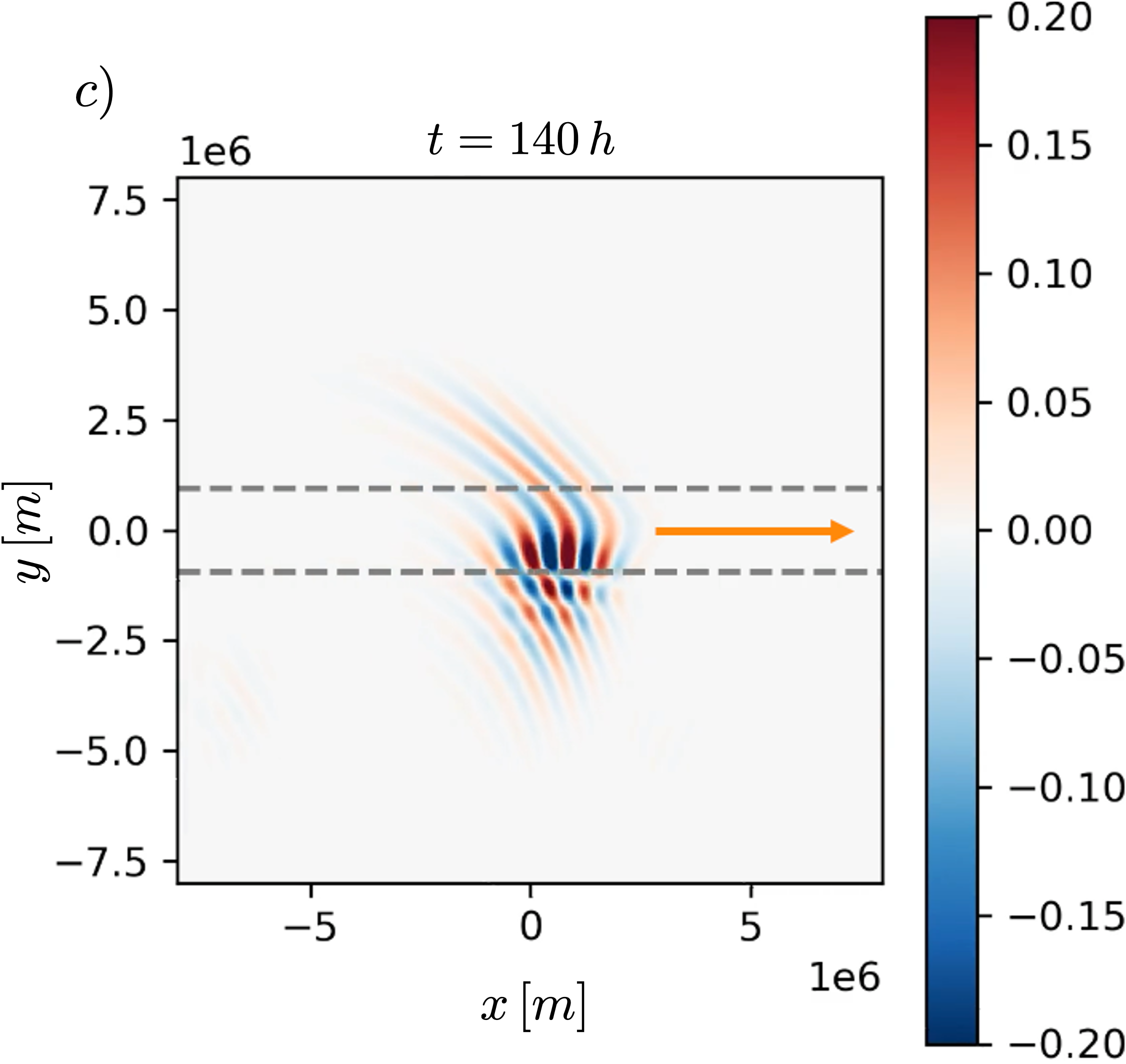}
    \includegraphics[width=0.465\linewidth]{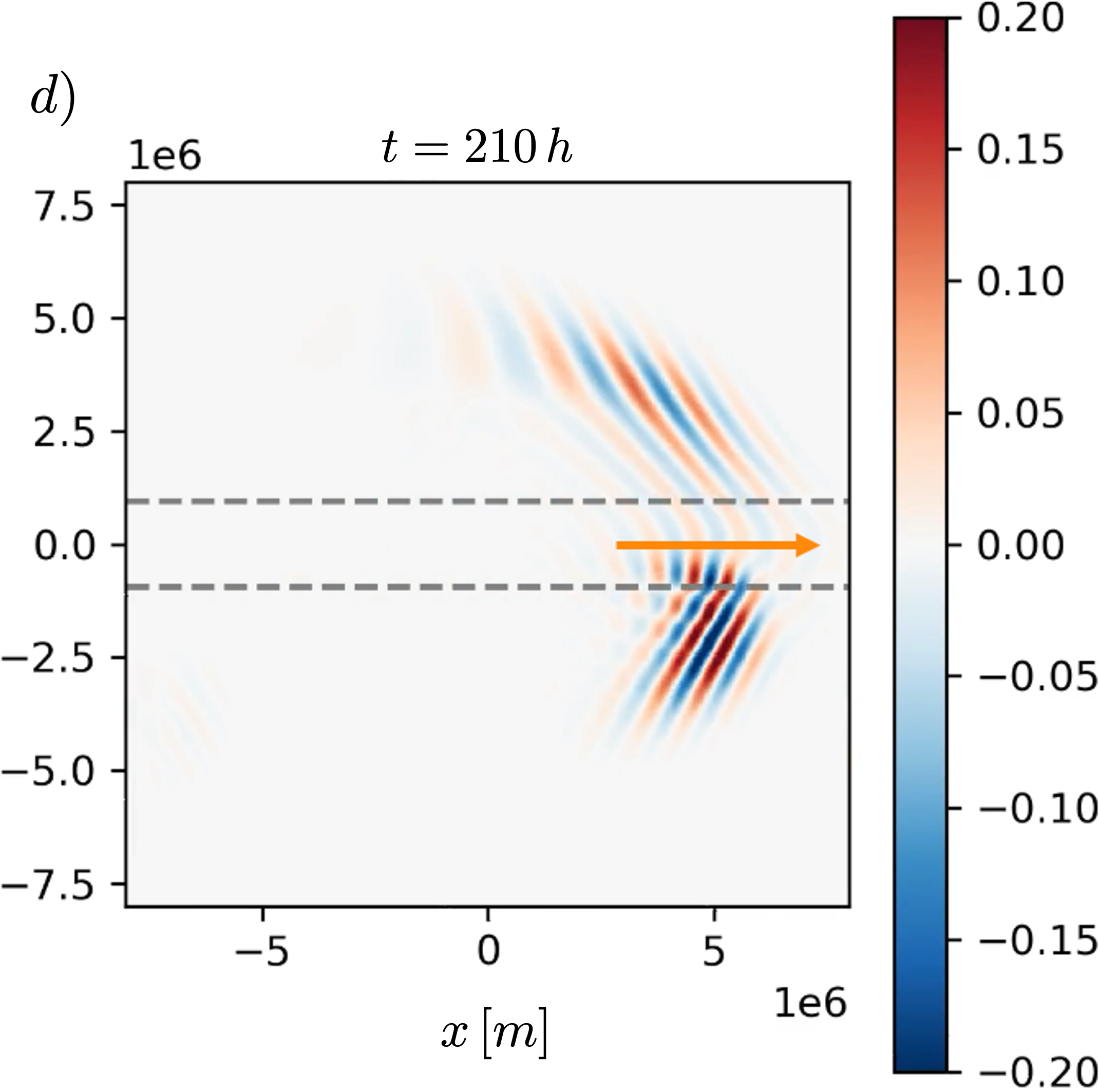}
    \captionsetup{width=1.\linewidth, justification=justified, format=plain}
    \caption{Zonal velocity (m$\cdot$s$^{-1}$) of vertical mode 1 M2 internal tide. The wavepacket is launched from $(x_0, y_0)=(-0.9\,L_x, -0.5\,L_y)$ with $\theta=7\pi/16$. Jet limits are denoted by horizontal grey dashed lines, its direction by the orange arrow. Jet is uniform on the vertical with U$_0=6$m$\cdot$s$^{-1}$.}
    \label{reflexion_uniformjet}
\end{figure}
\noindent Figures (\ref{critical_layer_uniformjet}) and (\ref{reflexion_uniformjet}) show two cases where the initial point of the wavepacket and its angle of launch ($(x_0,y_0)=(-7L_x/8,-L_y/2)$ and $\theta=7\pi/16$) have been optimized to maximize the interactions with the jet. The jet is also stronger in both cases to enhance the effect. For a jet of $U_0=-6$ m$\cdot$s$^{-1}$ (figure \ref{critical_layer_uniformjet}), the crests are strongly diverted in the direction of the current. When approaching this level, the increase in $k_{y}$ (smaller meridional wavelength) means the simulation under-resolves the waves. Conversely, for $U_0=6$ m$\cdot$s$^{-1}$ (figure \ref{reflexion_uniformjet}), the wavepacket largely reflects.

\newpage

\subsubsection{Energetics}
\begin{figure}
    \centering
    \includegraphics[width=\linewidth]{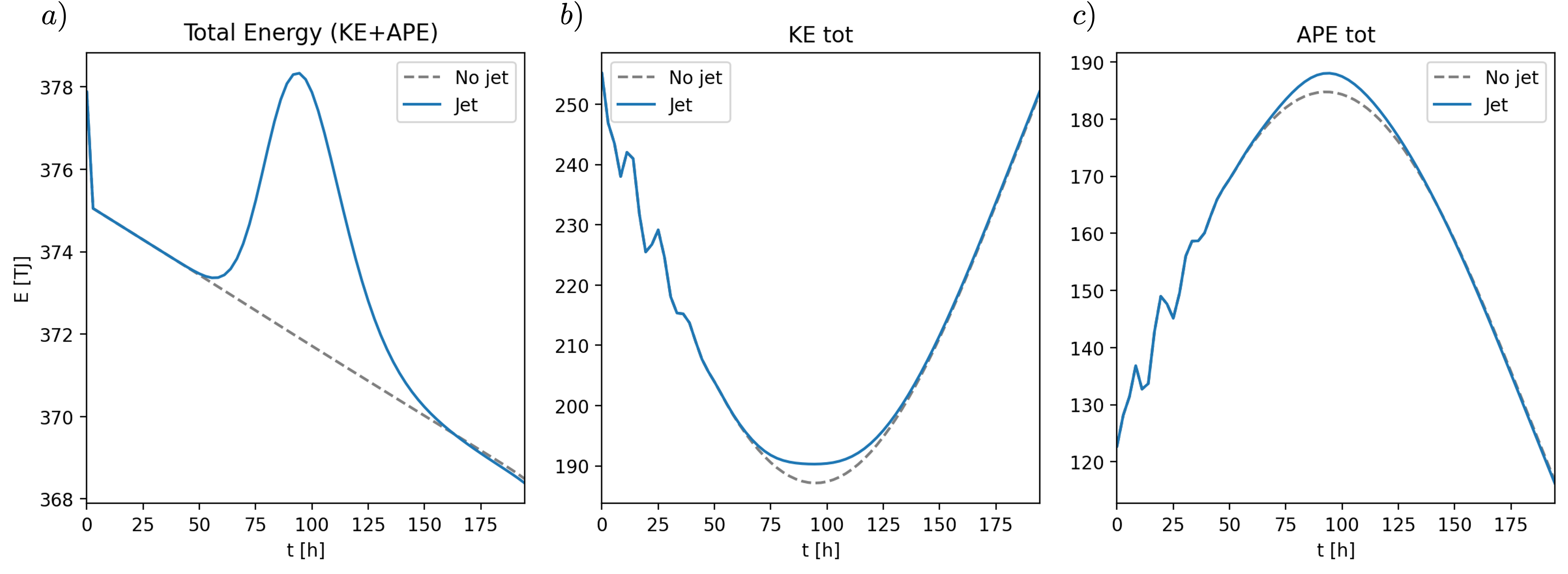}
    \captionsetup{width=1.\linewidth, justification=justified, format=plain}
    \caption{a) total energy, b) kinetic energy, c) available potential energy. Blue profiles correspond to the case with no jet, grey dashed lines for the jet uniform on the vertical with amplitude U$_0$=-1 m$\cdot$s$^{-1}$.}
    \label{energetics_uniform_jet}
\end{figure}
Figure (\ref{energetics_uniform_jet}) shows the total energy (TE; left panel), kinetic energy (KE; middle panel) and available potential energy (APE; right panel) of the mode 1 wavepacket with no jet as in figure (\ref{energetics_nojet}), also with a simulation computed with an westward jet having U$_0$=-1 m$\cdot$s$^{-1}$ (grey dashed profiles). Because in both cases the dynamics do not involve higher modes, the total wave energy is contained in mode 1.\\

The interaction between the jet and the wave causes the total energy of the wavepacket to rise significantly as it enters the jet (around $t\sim60$h), but the process reverses before the wavepacket exits the jet (around $t\sim160$h). This reversal of the energy transfer occurs after the wavepacket crosses the Equator at around $t\sim95$ h.\\

\noindent As no evolution of the jet or feedback from the jet is included into the model, the total energy is not conserved (in addition to the effect of dissipation, as shown in figure (\ref{energetics_uniform_jet}a)). In such configuration, it is the wave action $\mathcal{A}=$TE/$\omega_{\textrm{int}}$ that is conserved, provided that there is a scale separation between the wave and the jet.\\

\noindent Outside of the jet, where $\omega_{\textrm{int}}=\omega_{0}$, TE is given by:
\begin{equation}
	\textrm{TE}=\mathcal{A}\,\omega_{0},
\end{equation}
where $\omega_{0}$ is fixed at the excitation frequency of the M2 internal tide and $\mathcal{A}$ is affected by the dissipation, causing the initial decrease observed in figure (\ref{energetics_uniform_jet}a).\\
Inside the jet $\omega_{\textrm{int}}=\omega_{0}-E(y)k_{x0}$. With $E(y)<0$ as is the case in figure (\ref{energetics_uniform_jet}), $\omega_{\textrm{int}}$ increases to a maximum before decreasing back to  $\omega_{0}$ as the wavepacket exits the jet. As
\begin{equation}
	\textrm{TE}=\mathcal{A}(\omega_{0}-E(y)k_{x0}),
\end{equation}
it means that, superimposed to the steady decrease caused by dissipation, TE first increases to a maximum due to wave/jet interactions before decreasing back to the value imposed by the dissipation when the wavepacket exits the jet.

\newpage

\subsection{Vertical mode 2 Equatorial jet} \label{sect_shear_jet}
\subsubsection{Jet configuration and M2 dynamics}
The horizontal structure of the Equatorial jet is kept identical but its vertical structure corresponds to a vertical mode 2 ($n=2$): 
\begin{equation}
    U(y,z)=U_{0}\cos(m_{n}z)\textrm{exp}\left(-\frac{y^2}{2\rm{W}^2}\right),
    \label{sheared_jet_structure}
\end{equation}
where $m_{n}=2\pi/H$, $U_{0}=-1$ m$\cdot$s$^{-1}$, $\textrm{W}=400$ km and $H=5$ km.\\

\begin{figure}
 \centering
 \includegraphics[width=0.4\linewidth]{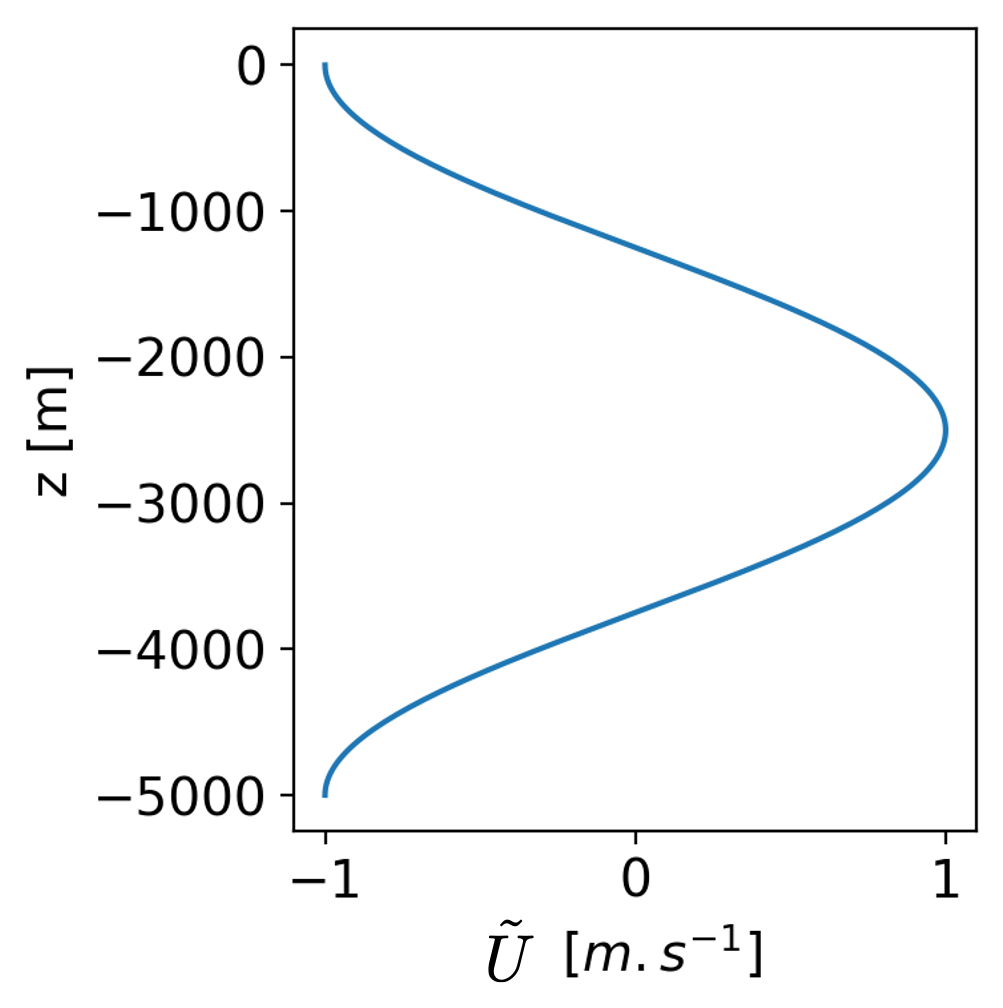}
  \caption{Vertical structure $\tilde{U}$ of the jet as described in (\ref{sheared_jet_structure}).}
  \label{jet_mode_2}
\end{figure}

\noindent The sheared vertical structure of the jet is displayed in figure (\ref{jet_mode_2}) and represents an idealised  Equatorial jet: westward at the top and bottom of the ocean but eastward in the center of the ocean.\\

\noindent With the jet sheared in the vertical, the model is driven by the modal equations (\ref{modal_eq_u}), (\ref{modal_eq_v}), (\ref{modal_eq_b}) with non-trivial interactions coefficients. Equations (\ref{C1C2}) and (\ref{C3C4C5}) also become non-trivial, representing the interactions between the vertical modes of the waves. In particular, for a jet with a vertical mode $2$ structure, the coefficients have non-zero values for the wave modes $m=n-2$ and $m=n+2$, except for the case $n=1$ which has also a non-zero interaction coefficient with itself ($m=n=1$).\\

\begin{figure}
    \centering
    \includegraphics[width=0.482\linewidth]{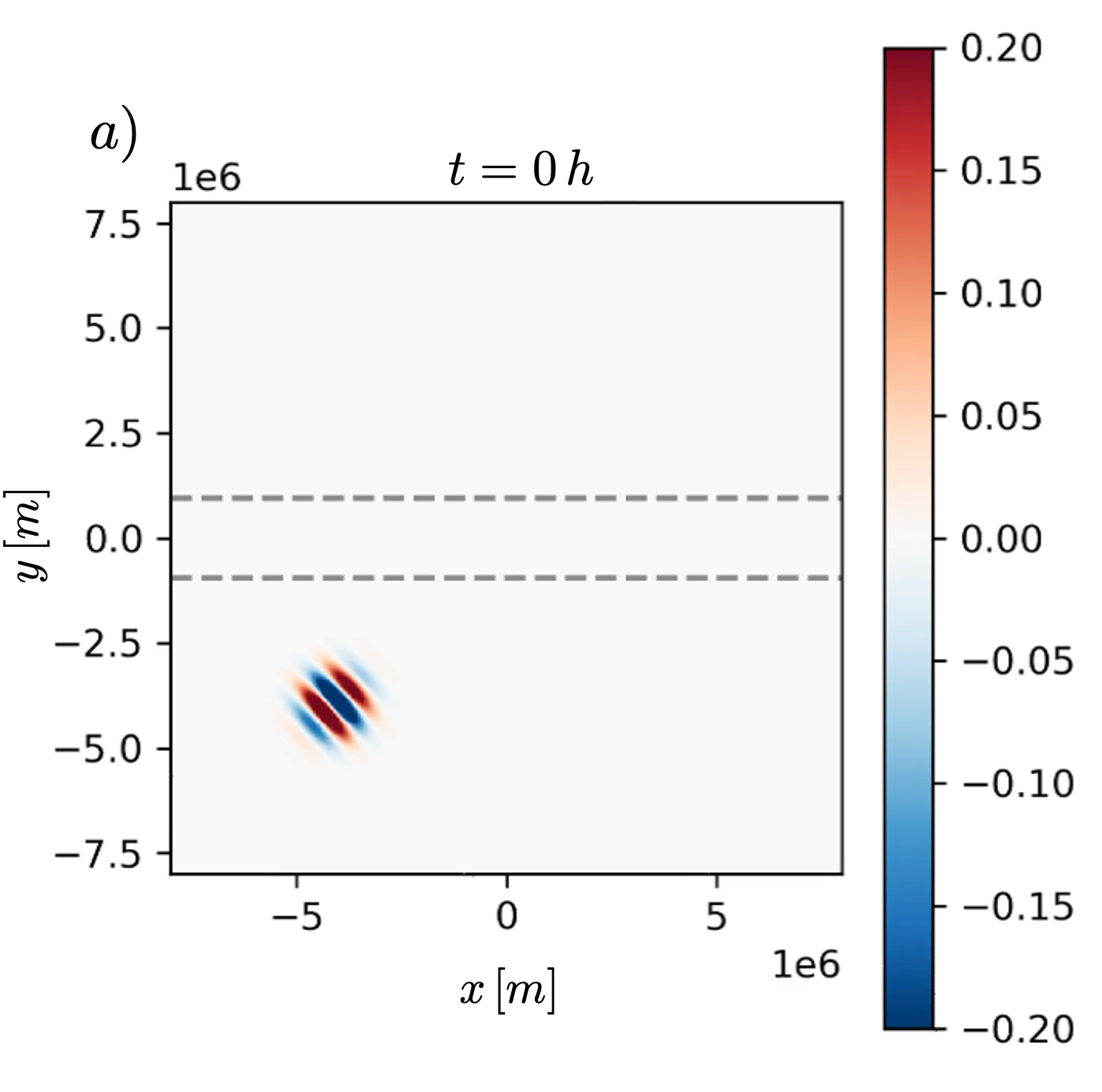}
    \includegraphics[width=0.468\linewidth]{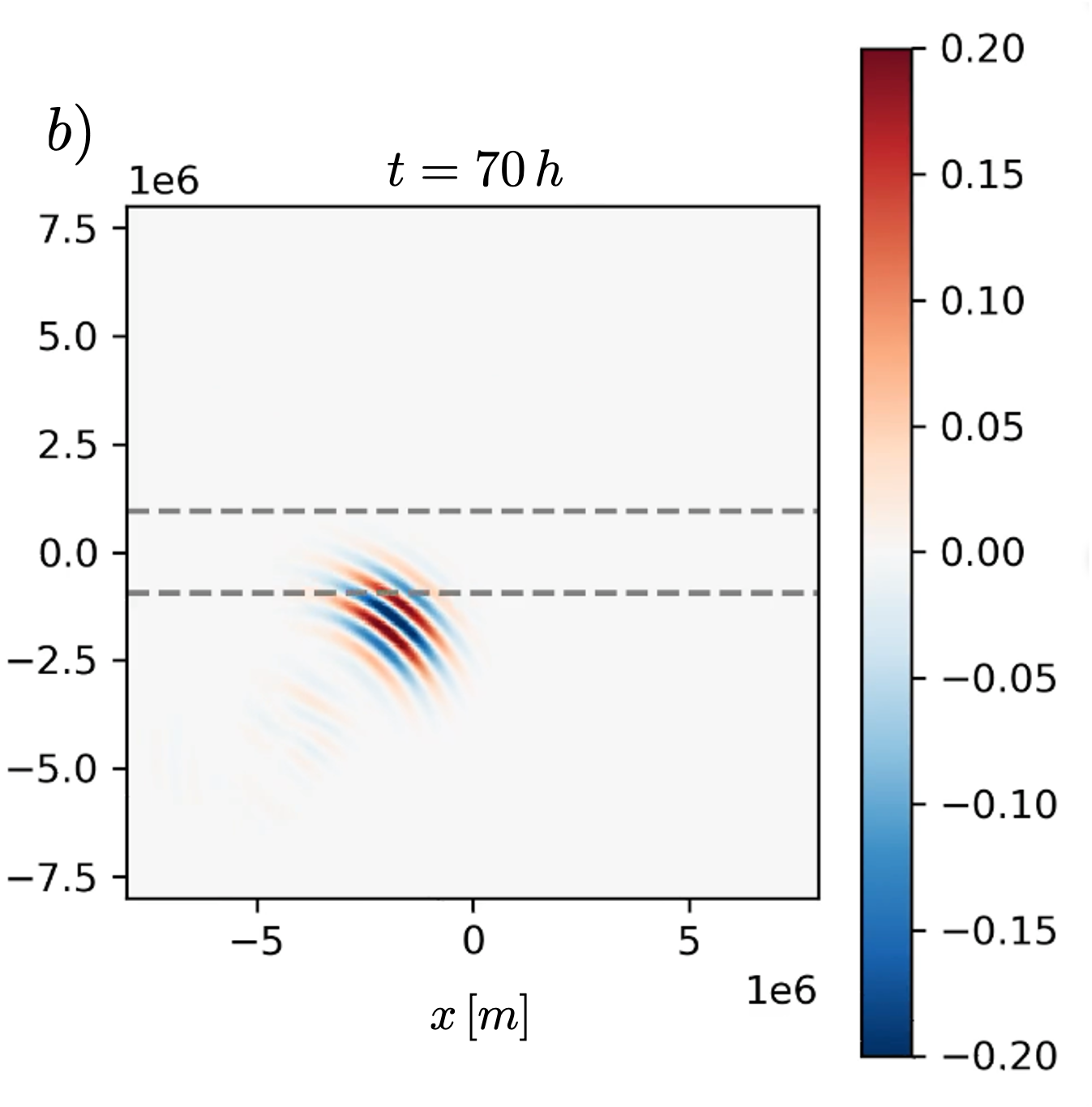}
    \includegraphics[width=0.482\linewidth]{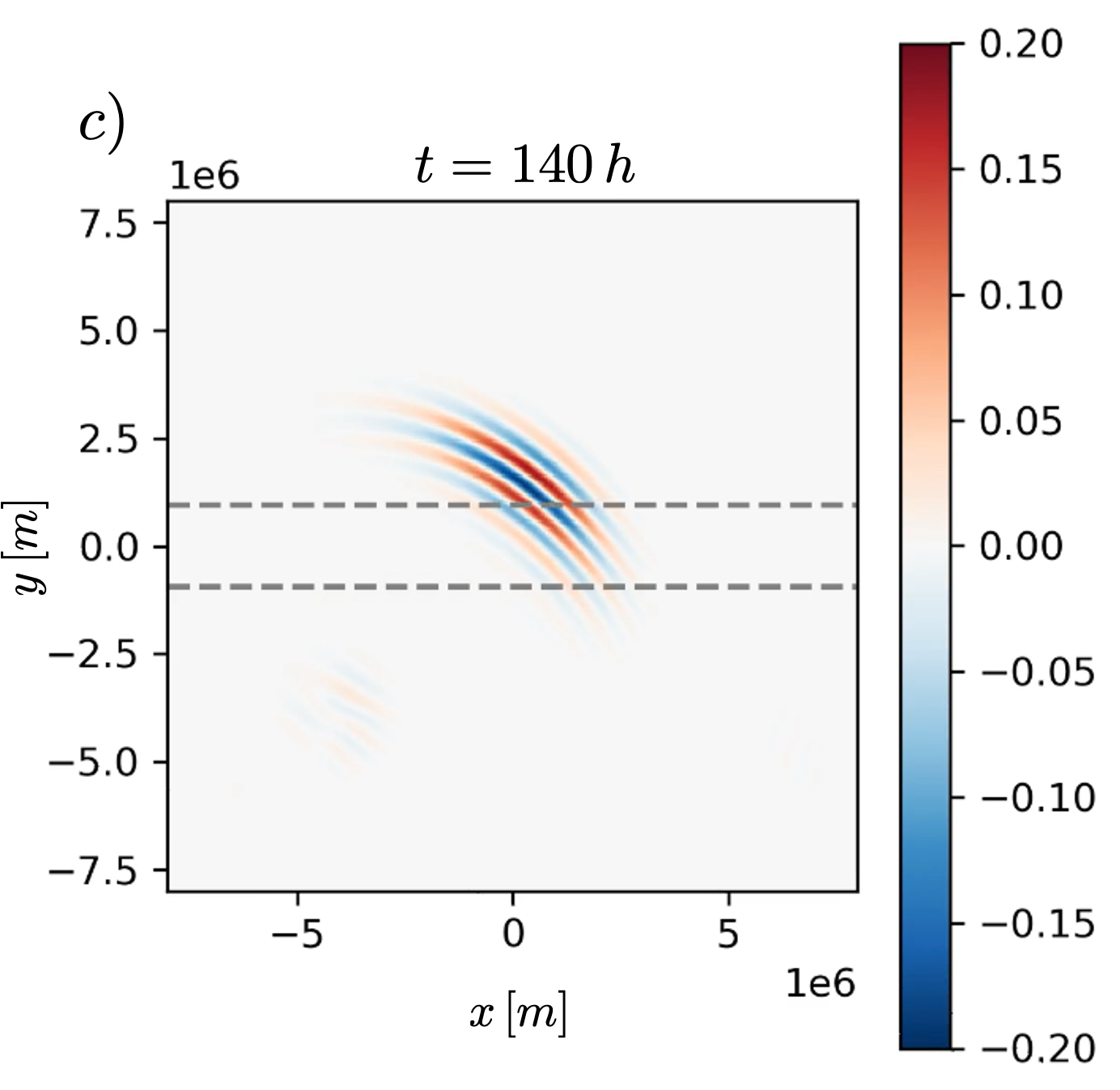}
    \includegraphics[width=0.468\linewidth]{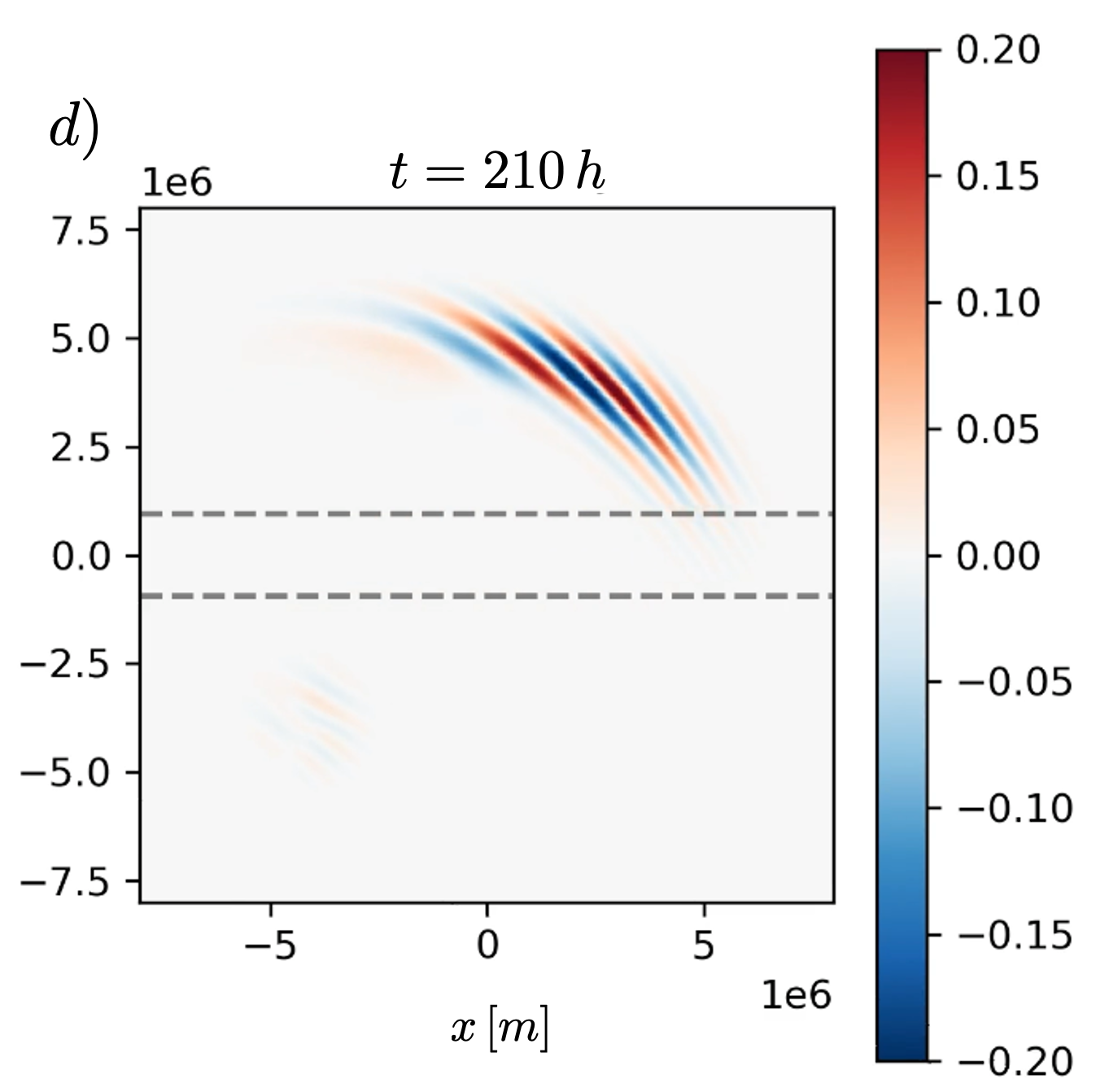}
    \captionsetup{width=1.\linewidth, justification=justified, format=plain}
    \caption{Zonal velocity (m$\cdot$s$^{-1}$) of vertical mode 1 M2 internal tide. The wavepacket is launched from $(x_0, y_0)=(-0.5\,L_x, -0.5\,L_y)$ with $\theta=\pi/4$. Jet limits are denoted by horizontal grey dashed lines, with a structure as in (\ref{sheared_jet_structure}).}
    \label{test_dynamic_shearedjet}
\end{figure}

\noindent Despite this, the evolution of the vertical mode 1 M2 internal tide is qualitatively similar to the cases with no jet or with a vertically uniform jet. For example, the evolution shown in figure (\ref{test_dynamic_shearedjet}) should be compared with figure (\ref{traj_raytracing_simulation_nojet}) that has no jet.\\

\begin{figure}
    \centering
    \includegraphics[width=0.86\linewidth]{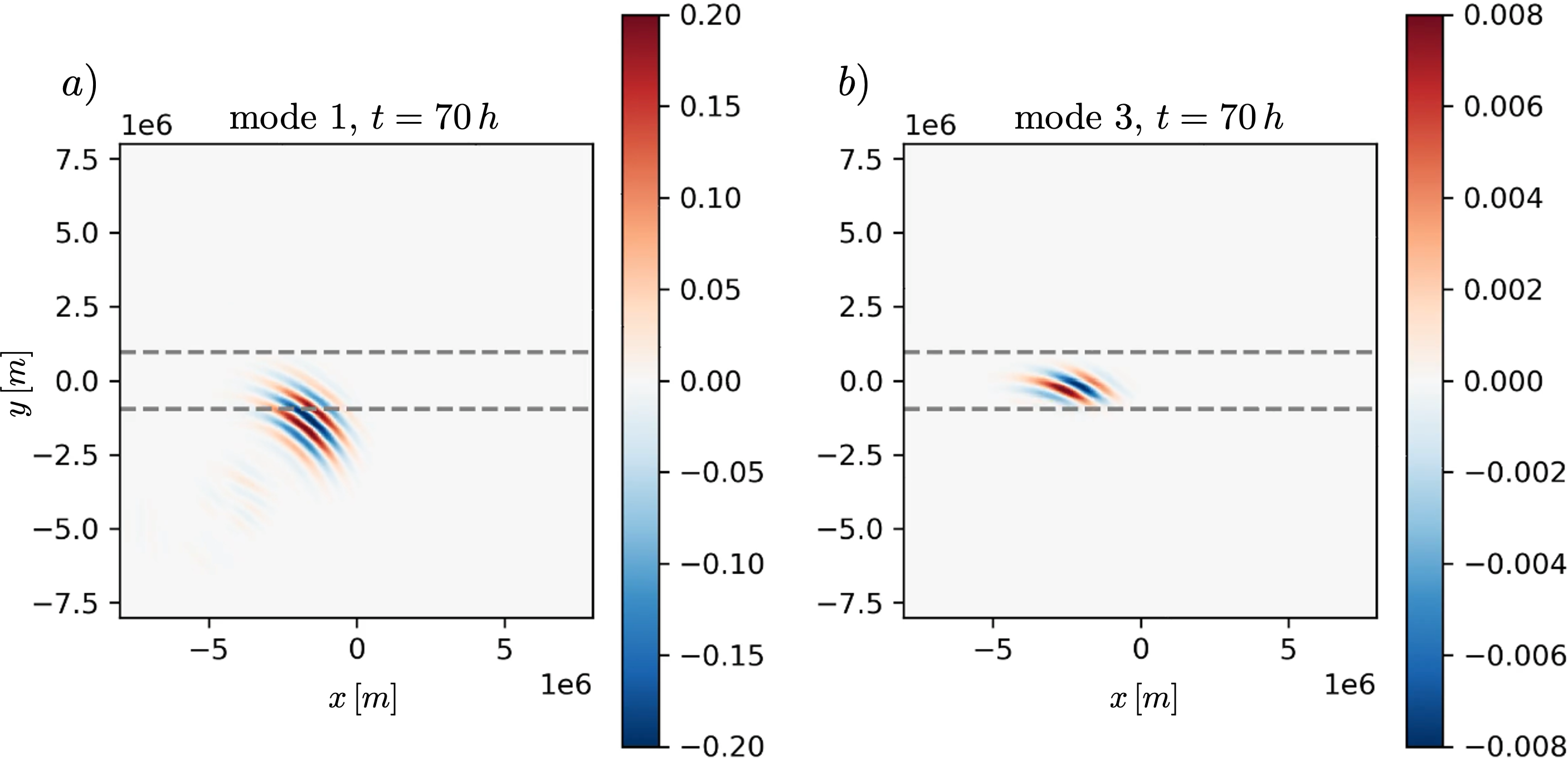}
    \includegraphics[width=0.86\linewidth]{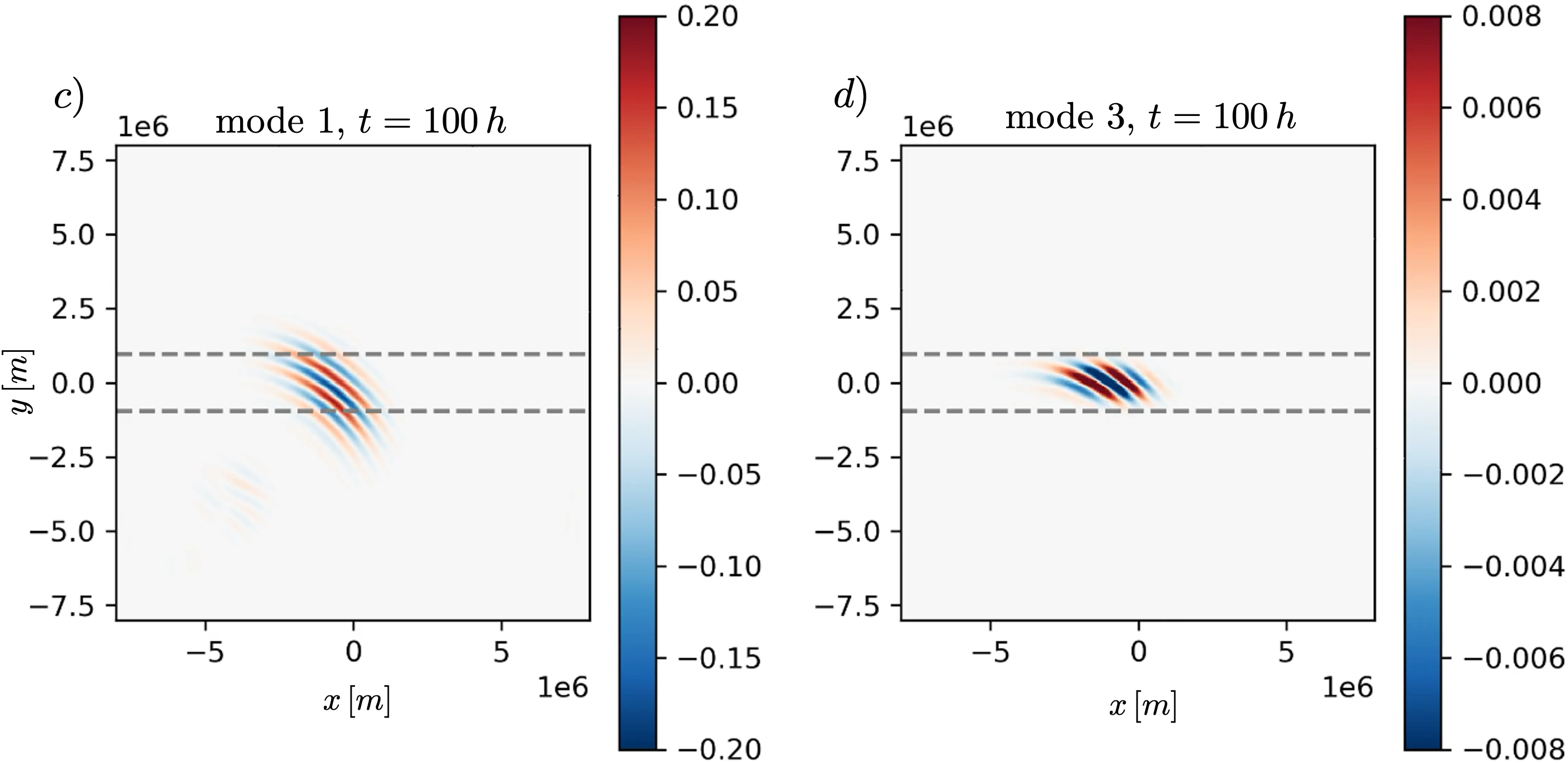}
    \includegraphics[width=0.86\linewidth]{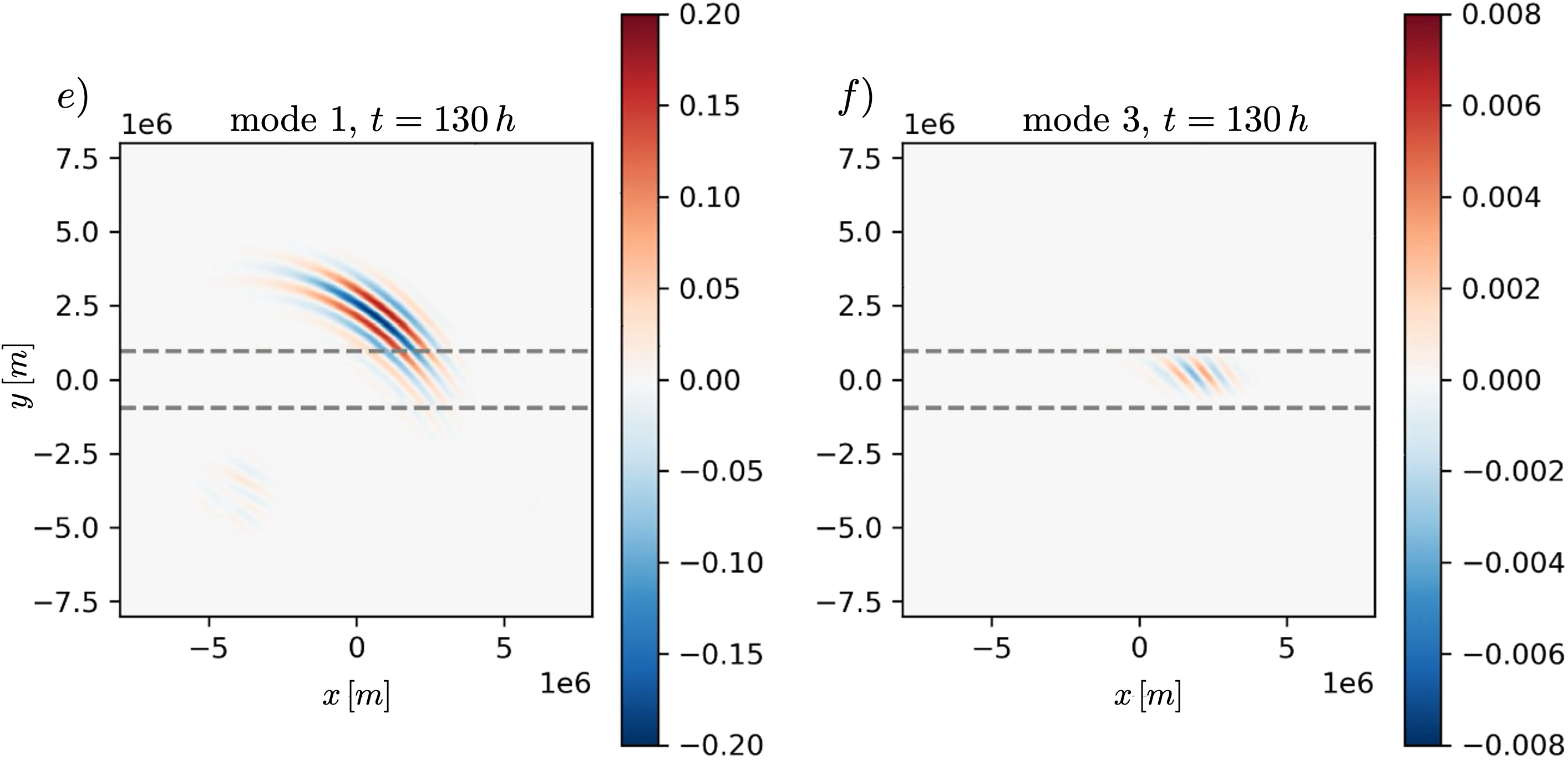}
    \captionsetup{width=1.\linewidth, justification=justified, format=plain}
    \caption{Zonal velocity (m$\cdot$s$^{-1}$) of vertical mode 1 (panels a), c), e)) and mode 3 (panels b), d), f)) M2 internal tide. The wavepacket is launched from $(x_0, y_0)=(-0.5\,L_x, -0.5\,L_y)$ with $\theta=\pi/4$. Jet limits are denoted in grey dashed lines, with a structure as in (\ref{sheared_jet_structure}).}
    \label{higher_modes_sheared_jet}
\end{figure}

\noindent Higher modes are excited, but at much smaller amplitudes than mode 1. Figure (\ref{higher_modes_sheared_jet}) shows how, the mode 1 wavepacket, when propagating inside the jet, generates mode 3 with smaller amplitude less than 10\% that of the mode 1 wave. This higher mode propagates inside the jet, in the same direction as the incident wave. It eventually decays, giving its energy back to mode 1 as the mode 1 leaves the jet. Such behaviour is also observed with higher modes, modes 5, 7, 9,..., all created inside the jet with smaller and smaller amplitudes and vanishing as the mode 1 wavepacket leaves the jet.\\

\noindent For the case treated here (where $N$ is constant) and near the equator where $\beta |y|\sim 0$, the phase and group speeds are inversely proportional to the mode number. As a result, the wavepacket envelope as well as the crests of higher modes are slower than those of the initial mode 1. For example, for $\theta=\pi/4$, the mode 1 velocities are $c_{p1}=27$m$\cdot$s$^{-1}$ and $c_{g1}=9$m$\cdot$s$^{-1}$, whereas the mode 3 has $c_{p3}=9$m$\cdot$s$^{-1}$ and $c_{g3}=3$m$\cdot$s$^{-1}$.\\

\noindent These higher modes also have smaller horizontal wavelengths (the wavenumbers grow with the mode number), which would cause then to be more susceptible to dissipation and encountering critical layers. Numerically it means that the number of modes solved for is limited (to 10 in our case) as well as the maximum velocity of the jet.

\subsubsection{Energetics}
\begin{figure}
    \centering
    \includegraphics[width=\linewidth]{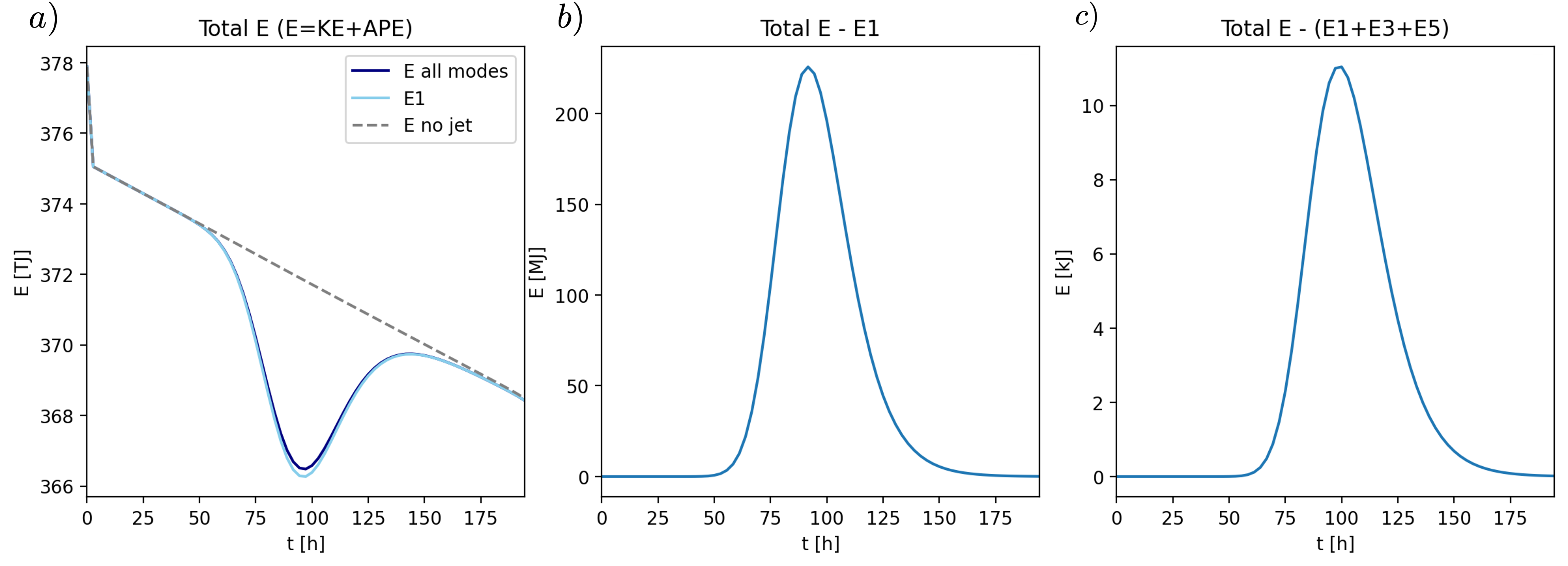}
    \captionsetup{width=1.\linewidth, justification=justified, format=plain}
    \caption{a) total energy. Grey dashed profile corresponds to the case with no jet, light blue to the total energy of mode 1 with the sheared jet and darker blue to the total energy of all the modes combined, in the case of the sheared jet. b) difference between the total energy in the domain and the total energy of the mode 1. c) difference between the total energy in the domain and the total energy of the modes 1, 3 and 5 combined.}
    \label{energetic1_sheared_jet}
\end{figure}

Figure (\ref{energetic1_sheared_jet}) shows the total energy of the wavepacket (left panel) for the case with the mode 2 jet, first for all the vertical modes (dark blue profile), then for the mode 1 only (light blue profile). The case with no jet has also been added (grey dashed line). Compared to the variation caused by the wave-mean flow interaction, the difference brought by the creation of the higher modes is small. However, looking at the middle panel figure (\ref{energetic1_sheared_jet}) that represents the difference between the total energy of all the modes and that of mode 1, shows that this difference still accounts for more than 200 MJ and develops only when the M2 internal waves crosses the jet. In the right panel of figure (\ref{energetic1_sheared_jet}), this difference in the total energy is mostly explained when also adding the total energies of mode 3 and mode 5. As such, it shows how the higher modes created inside the jet are the result of the scattering of the energy from mode 1 to the modes of higher order, made possible by the interaction with the jet's non-uniform vertical structure. \\

The difference between the total energy and the energy of the mode 1, due to the scattering of energy to higher modes, is observed to decay at later times, when the wavepacket exits the jet, it corresponds to the vanishing of the higher modes signal observed in figure (\ref{higher_modes_sheared_jet}). This scattered energy is returned to the mode 1, as observed for the energy of the wave-mean flow interaction in figure (\ref{energetics_uniform_jet}). Also, successive higher modes created receive less and less energy, as their amplitudes decrease with mode number.\\

\begin{figure}
    \centering
    \includegraphics[width=\linewidth]{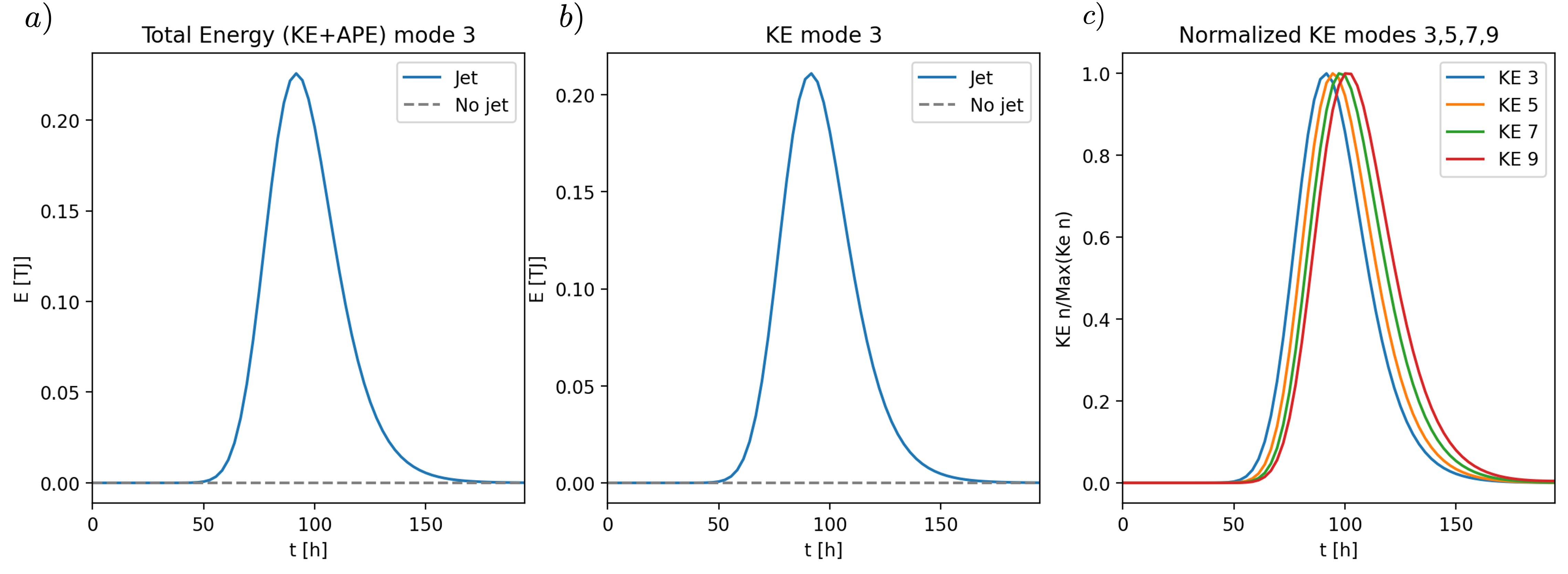}
    \captionsetup{width=1.\linewidth, justification=justified, format=plain}
    \caption{a) total energy of mode 3. Grey dashed profile corresponds to the case with no jet and blue profile to the sheared jet case. b) kinetic energy of mode 3. c) kinetic energy of modes 3 (blue), 5 (orange), 7 (green) and 9 (red), normalized by their maximum values respectively.}
    \label{energetic2_sheared_jet}
\end{figure}

\noindent The left panel of figure (\ref{energetic2_sheared_jet}) shows the total energy of the mode three, which is the first to be created through wave-mean flow interactions and also the higher mode that receives the most energy. The comparison with the case without the jet is added in grey dashed lines to show how the creation is only possible with the addition of the vertically mode 2 jet. The middle panel of figure (\ref{energetic2_sheared_jet}) presents only the kinetic energy of the mode 3. It demonstrates how the scattering of energy to the higher modes is realised through the growth of kinetic energy as it nearly accounts for the all of the mode's energy in comparison to the previous panel. The right panel of figure (\ref{energetic2_sheared_jet}) shows the kinetic energy of all the higher modes created (the maximum number of vertical modes solved for here is 10), normalised by their maximum value. It shows how the modes are created subsequently, mode 3 then mode 5, 7 and 9. The reverse transfer of energy toward mode 1 then occurs in the inverse order, mode 9 being the first to give all its energy back to mode 7 then mode 7 to mode 5 and so forth to mode 1. This effect is somewhat masked by the normalisation (right panel of figure (\ref{energetic2_sheared_jet})) as it shows that mode 9 seems to be the last to vanish. However, mode 9 represents so little energy compared to mode 3, that the tail of the distribution of mode 3 (around $t\sim 160$) still contains more of the total energy than what is distributed amongst the higher modes. The last mode to transfer its energy to mode 1 is then effectively mode 3.\\


\begin{figure}
    \centering
    \includegraphics[width=0.42\linewidth]{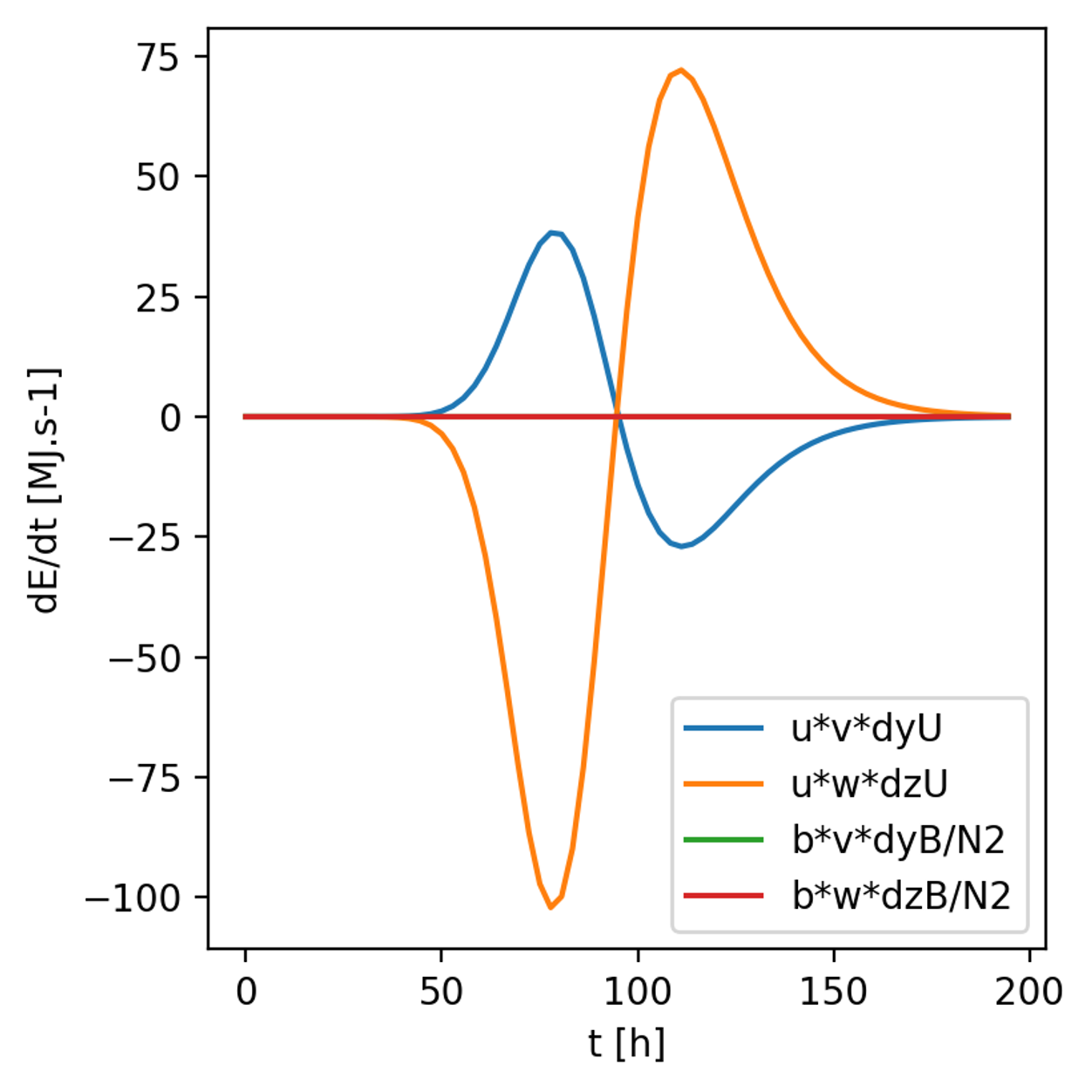}
    \captionsetup{width=1.\linewidth, justification=justified, format=plain}
    \caption{Variation of the different terms of the energy budget, as computed in (\ref{bilan_energie}).}
    \label{energetic3_sheared_jet}
\end{figure}

Figure (\ref{energetic3_sheared_jet}) shows the evolution of each term of the energy budget (\ref{bilan_energie}). The dominant terms are those of shear production (blue and orange profiles), related to the horizontal and vertical shear of the mean flow. Both shear contributions change sign after the wavepacket crosses the Equator. The part due to the vertical shear of the mean flow constitutes the biggest contribution to the energy budget as the shear is more intense being localised over the 5 km depth, as opposed to the 400 km of width the Equatorial jet. The terms of buoyancy production do not contribute to the evolution of the total energy.

\subsubsection{Possible incoherence?}
Coming back to the motivating question, we want to check if the wave-mean flow interaction could explain the lack of observed M2 internal tide signal in the Equatorial Pacific \citep[][]{Buijsmanetal2017}.\\

We found the overall propagation of the tidal mode 1 negligibly changes with or without a sheared jet (figure (\ref{test_dynamic_nojet}) and (\ref{test_dynamic_shearedjet})) and the processes related to the excitation of higher modes reverse when the wavepacket exits the jet (see section \ref{sect_shear_jet}). To check if there are any differences due to the interaction with and without a jet, we assess wave incoherence, from the difference between the case with the mode 2 jet and the case with no jet.\\

\begin{figure}
    \centering
    \includegraphics[width=0.405\linewidth]{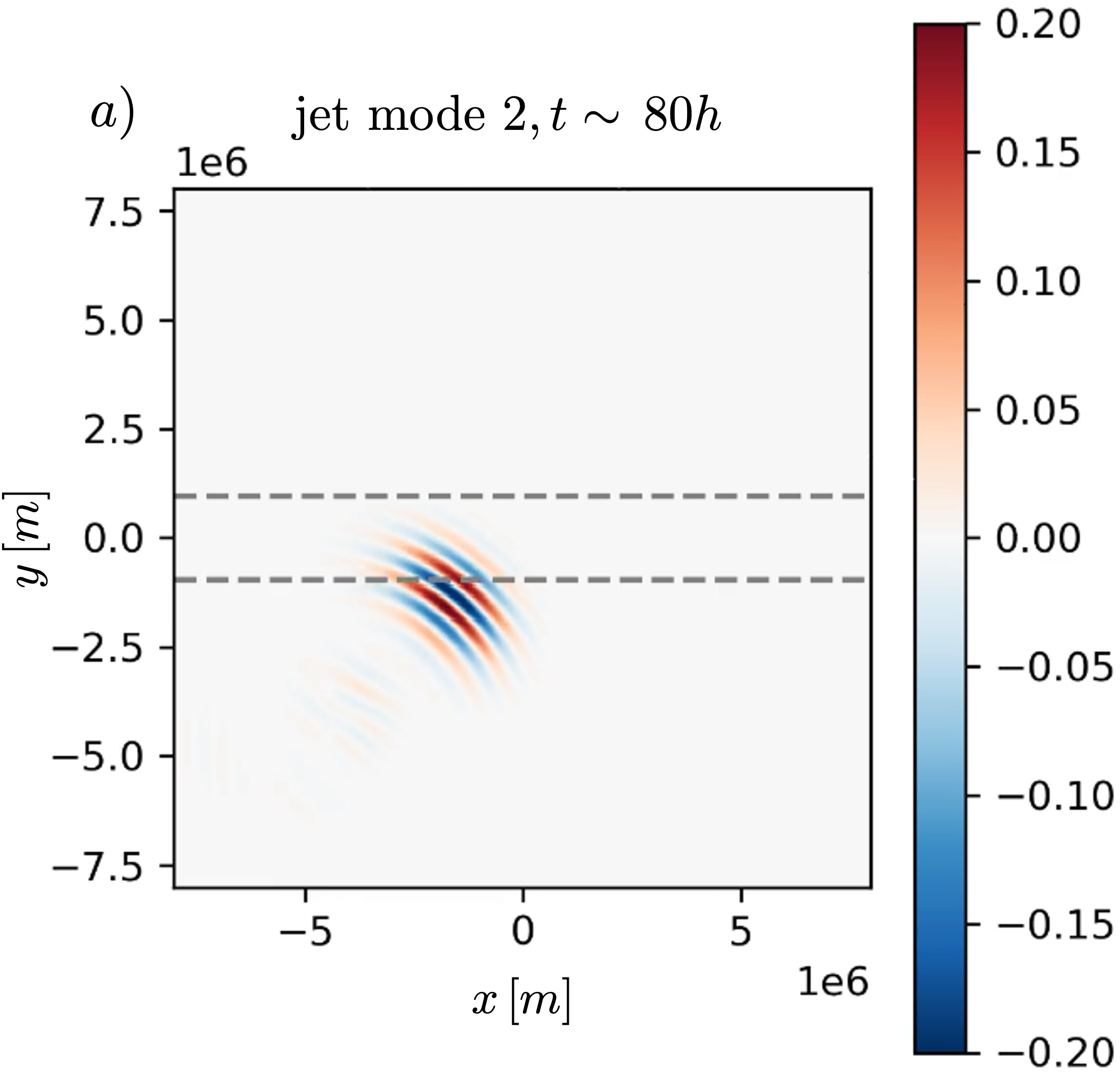}
    \includegraphics[width=0.4\linewidth]{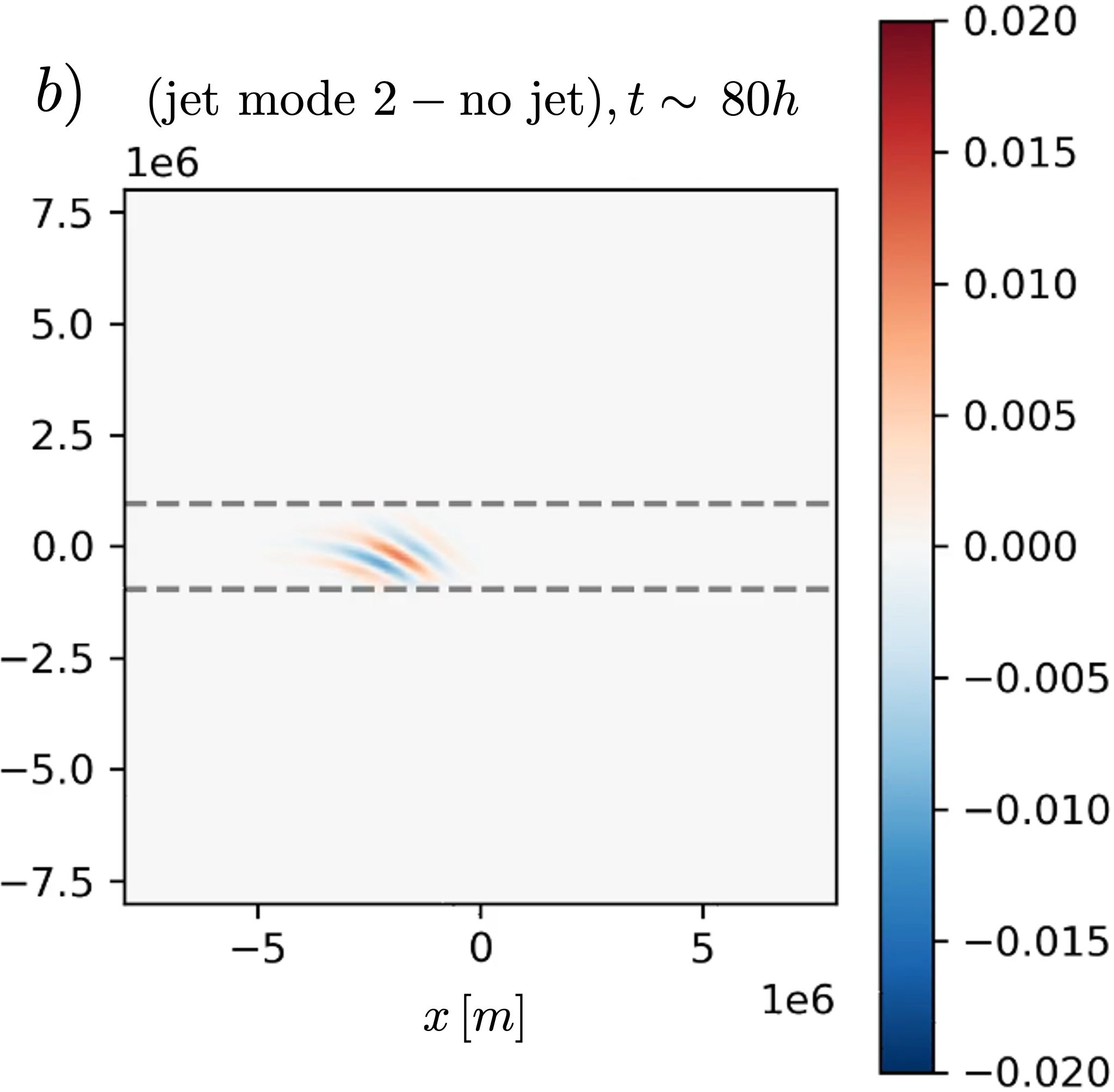}
    \includegraphics[width=0.405\linewidth]{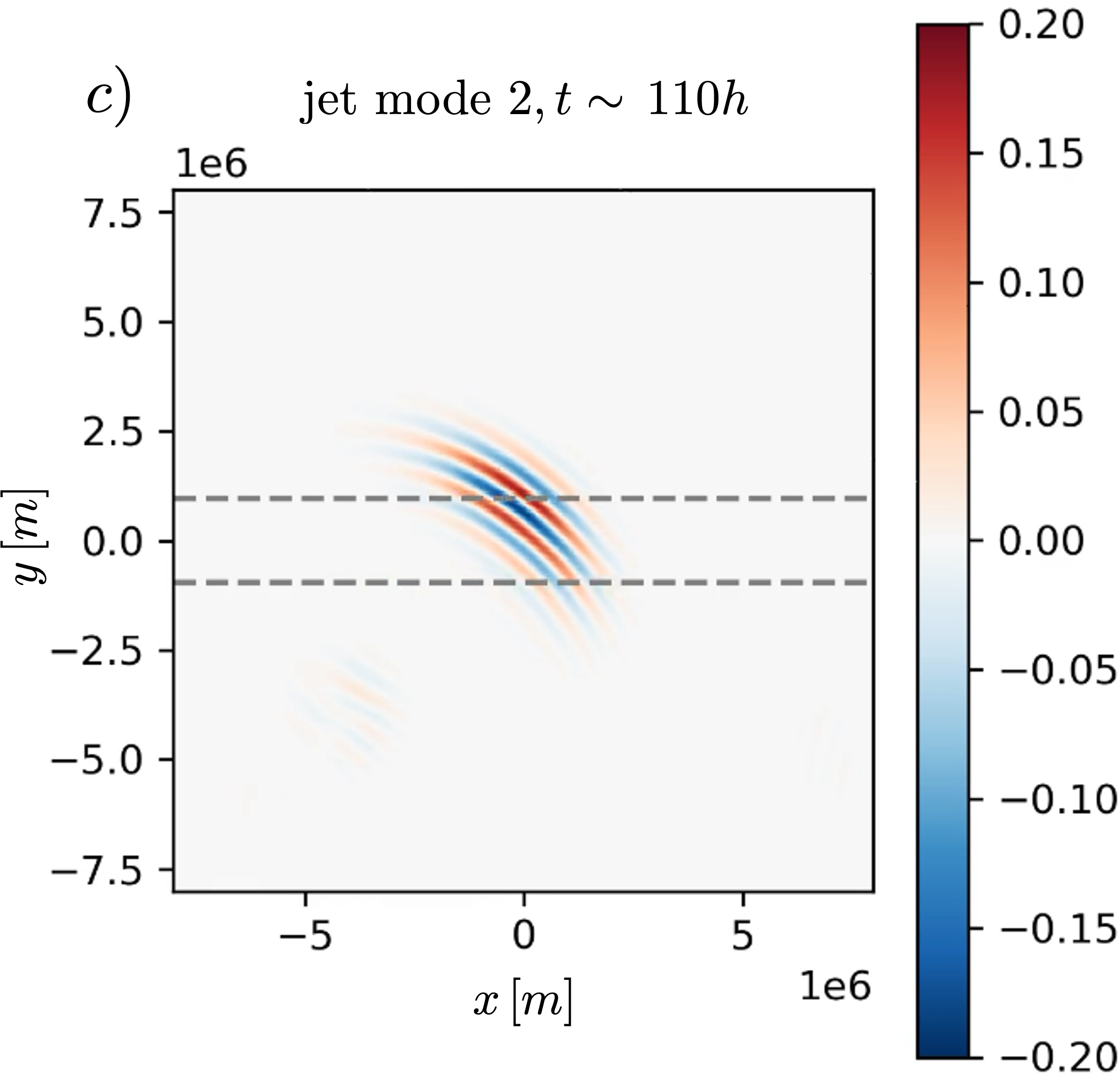}
    \includegraphics[width=0.4\linewidth]{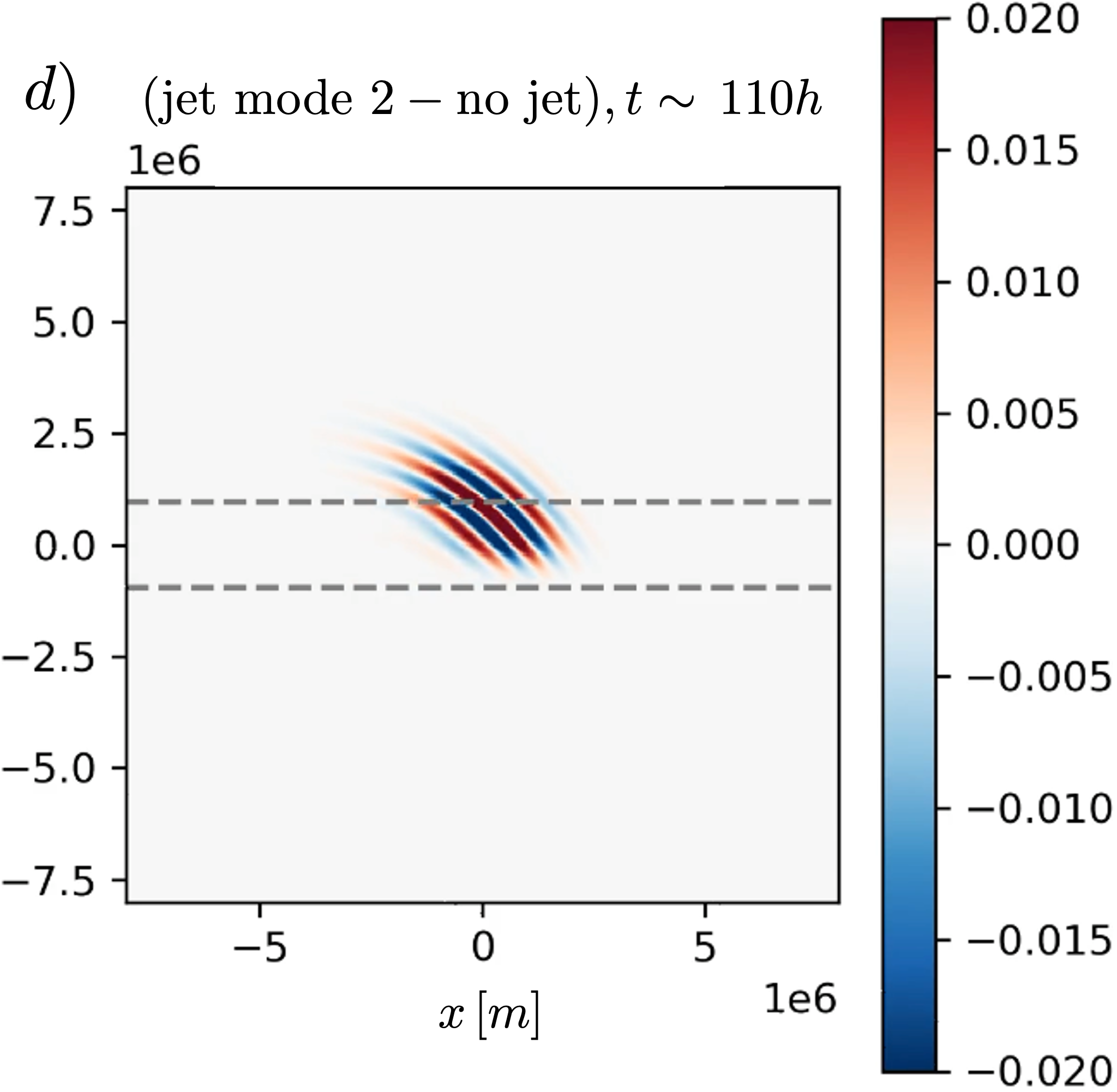}
    \includegraphics[width=0.405\linewidth]{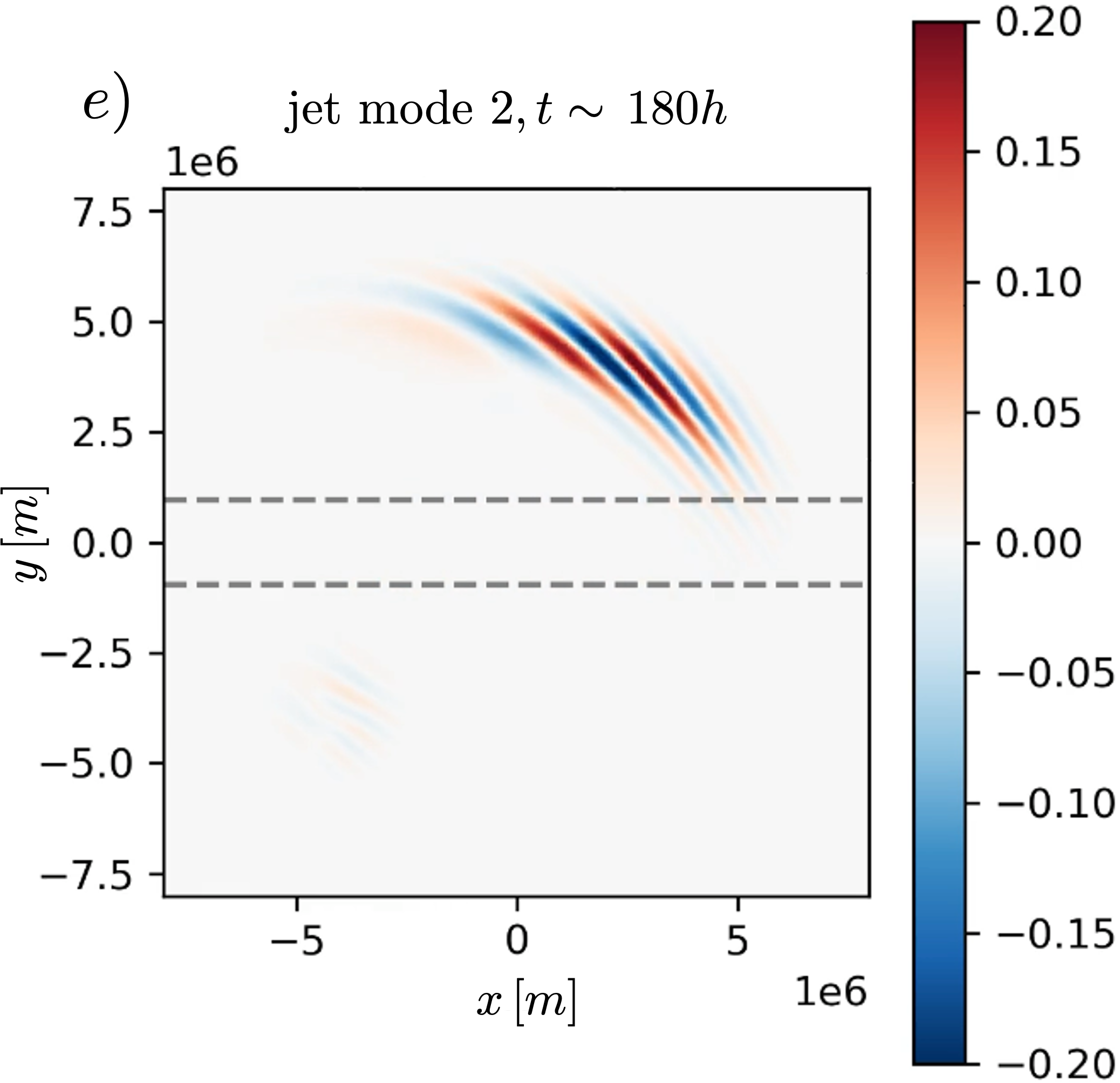}
    \includegraphics[width=0.4\linewidth]{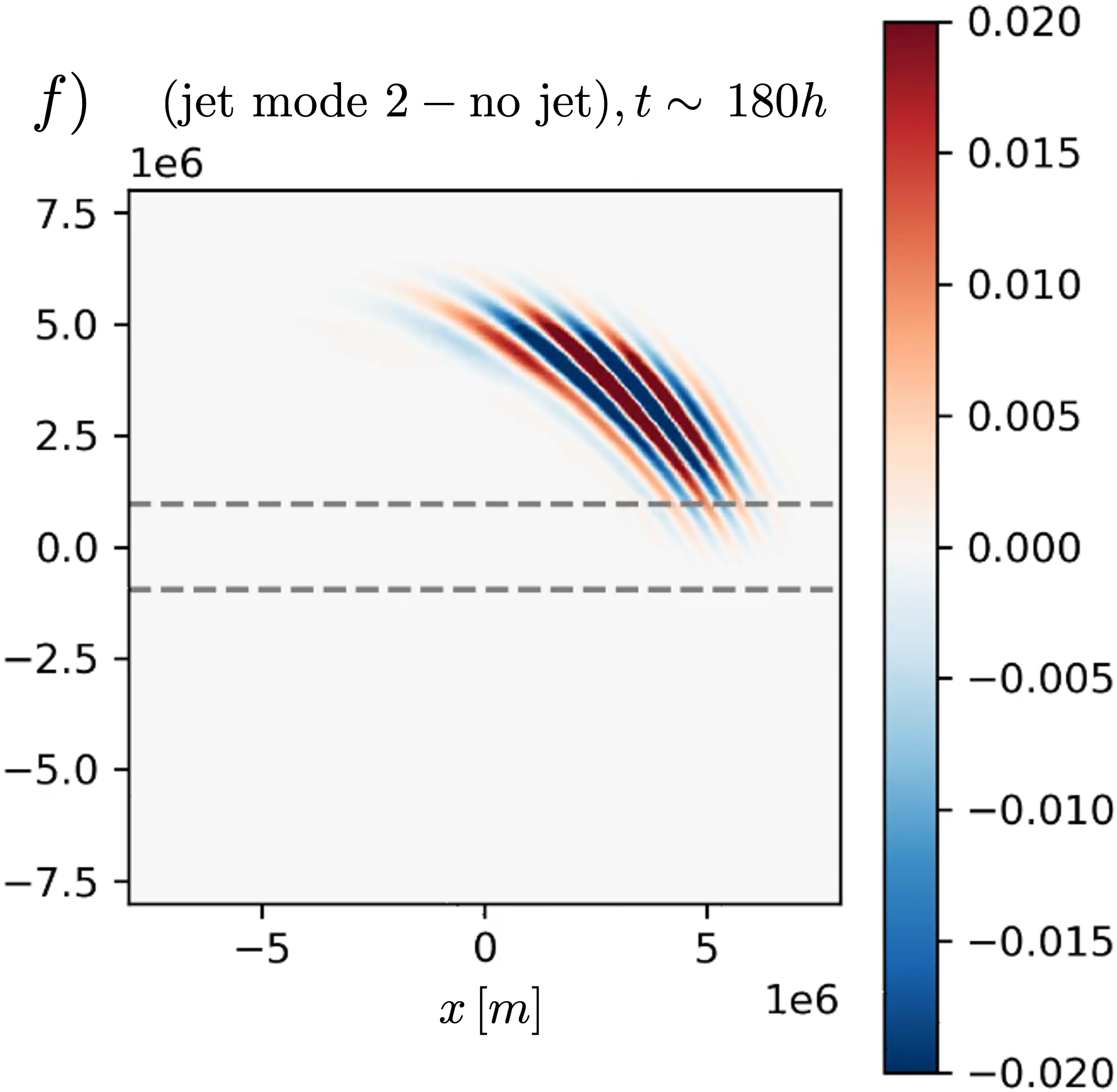}
    \captionsetup{width=1.\linewidth, justification=justified, format=plain}
    \caption{Zonal velocity (m$\cdot$s$^{-1}$) of vertical mode 1 M2 internal tide with the sheared mode 2 jet (panels a), c), e)). Difference in the zonal velocity of vertical mode 1 M2 internal tide between the case with the sheared jet and the case with no jet (panels b), d), f)). The wavepacket is launched from $(x_0, y_0)=(-0.5\,L_x, -0.5\,L_y)$ with $\theta=\pi/4$. Jet limits are denoted in grey dashed lines, with a structure as in (\ref{sheared_jet_structure}).}
    \label{simu_incoherence}
\end{figure}
\begin{figure}
    \centering
    \includegraphics[angle=90,width=0.325\linewidth]{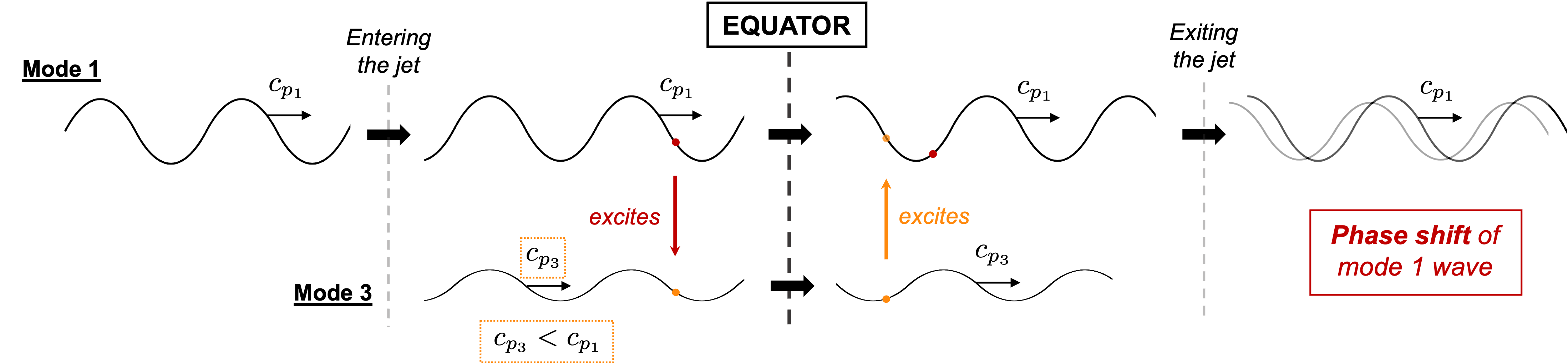}
    \captionsetup{width=1.\linewidth, justification=justified, format=plain}
    \caption{Scheme of the incoherence mechanisms due to wave-mean flow interaction and the creation of higher modes.}
    \label{schema_incoherence}
\end{figure}
Figure (\ref{simu_incoherence}) shows how a difference in the dynamics is indeed created by the addition of the sheared jet. The difference only becomes significant when the wavepacket enters the jet and grows as long as both interact. The difference remains after the wavepacket is well north of the jet. This is unlike the dynamics and energy transfer observed for the creation of higher modes and it points toward the creation of lasting incoherence due to the wave-mean flow interaction.\\

Figure (\ref{schema_incoherence}) is a schematic representation of our interpretation of such incoherence. As the mode 1 wavepacket enters the vertically sheared jet, energy is scattered toward higher modes. These modes propagate at a speed that is lower than that of mode 1 (both the group and phase speed are lower the higher the mode number). When the reverse transfer of energy occurs as the wavepacket propagates north of the Equator, the energy is transferred back to mode 1, but with a shift caused by the difference of phase speeds. The mode 1 wavepacket thus becomes phase shifted, creating incoherence in the internal tide signal. This incoherence is a possible explanation of the lack of internal tide signal detected north of the equator in the Equatorial Pacific [1]. %\citep[][]{Buijsmanetal2017}.

\newpage

\section{Conclusion}\label{conclusion}
We have investigated, by means of theory and numerical simulations within a idealised setup, the wave-mean flow interaction between a mode 1 M2 internal tide and a vertically uniform and sheared equatorial jet. The aim was to validate if such interactions could result in incoherence and hence explain the lack of M2 altimetry signal in the Equatorial Pacific.\\

We have identified different effects that can influence the dynamics of the M2 wavepacket. The $\beta$-effect deflects the path of the wavepacket but has negligible influence immediately at the Equator. A vertically uniform jet could, depending on its strength, lead to total reflection of the wavepacket or strong distortion of the crests. A vertically sheared jet however scatters energy to higher modes. These higher modes have smaller horizontal wavelength but also lower phase and group speed. The transfer of energy is reversed after the wavepacket crosses the Equator, the higher modes vanishing and all the energy going back into mode 1. It is the difference of propagation speed between the different modes that could explain the creation of incoherence in the signal, that would affect the altimetry observations [1].\\

Because of their smaller horizontal wavelength and lower phase and group speed, the higher modes excited by the jet are likely to be more sensitive to dissipation, but also to critical layers in a case where $N$ is not constant. It could cause further deposition of energy inside the jet and hence modify its dynamics as well as that of the wavepacket. It is an effect that we will develop further in future work.\\

The theoretical tools developed here are readily adapted to realistic profiles of stratification and equatorial currents. In particular, future work will model stratification motivated by observations and a zonal equatorial jet with strong westward near-surface flow and underlying countercurrents, as shown in figure (\ref{presentation_equatorial_jet}).

\section{Acknowledgements}
I sincerely thank Bruce Sutherland and Lois Baker who supervised this project and from whom I learned a lot during this summer. They showed great patience and unwavering support at every step of the way, for which I am extremely grateful. I also want to deeply thank Keaton Burn and Eric Ester for the precious help they provided regarding Dedalus and how to build a numerical model from scratch. At last, I wish to profoundly thank everyone that was at Walsh Cottage this summer and who made this experience so wonderful and enriching.
%you can add an acknowledgement if you wish. This is where you would mention your advisor(s).

\end{document}